\documentstyle[epsf,aps]{revtex}
\begin{document}

\draft
\title{Nucleon self-energy in the relativistic Brueckner approach}
\author{L. Sehn, C. Fuchs, Amand Faessler }
\address{Institut f\"ur Theoretische Physik, Auf der Morgenstelle 14,
        72076 T\"ubingen }
\maketitle

\begin{abstract}
The formalism of the relativistic (or Dirac-) Brueckner approach
in infinite nuclear matter is described.
As nucleon-nucleon interaction the one-boson exchange 
potentials Bonn A,B,C and for comparison the Walecka model are used.
The T-matrix is determined from the Thompson equation 
and is projected onto five covariant amplitudes.
By the restriction to positive energy states an ambiguity arises
in the relativistic Brueckner approach which is discussed here in terms 
of the pseudo-scalar and the pseudo-vector projection.
The influence of the coupling of the nucleon via the T-matrix as an effective
two-nucleon interaction to the nuclear medium 
is expressed by the self-energy.
In particular we investigate the scalar and vector components of the 
self-energy for the different  one-boson exchange potentials and 
discuss their density and momentum dependence.
We estimate the uncertainty of the self-energy due to the
pseudo-scalar and the pseudo-vector choice.
Usually the momentum dependence of the self-energy is thought to be weak,
however, we find that this depends on the one-boson exchange 
potentials.
For the Bonn potentials, in contrast to the $\sigma\omega$-potential, 
the momentum dependence is strikingly strong 
above as well as below the Fermi surface. 
We compare with the results of other groups and study
the effects on the equation of state and the nucleon optical potential.
\end{abstract}

%%         revtex version: pacs no. 
\pacs{21.30.+y, 21.65.+f, 24.10.Cn}

\section{Introduction}
In the relativistic (or Dirac-) Brueckner model
the dynamical two-nucleon correlations in infinite nuclear matter are studied.
The formalism  is based on an effective quantum field theory for mesons 
and nucleons \cite{3} and thus ignores the underlying quark nature of 
the nucleon. In the Brueckner model the T-matrix (or Brueckner 
G-matrix) serves  
as an effective in-medium two-body interaction. It is
determined by a self-consistent summation of the ladder diagrams in a 
quasipotential approximation (Thompson equation) to the Bethe-Salpeter 
equation, see, e.g., Ref. \cite{15}. The bare nucleon-nucleon (ladder) 
interaction is
described by one-boson exchange potentials. 
In the present work we apply the Bonn potentials A,B,C of Refs. \cite{4,5}
and for comparison also a simple $\sigma\omega$-potential which is well 
known from the Walecka model \cite{3}.
The nucleon inside the nuclear medium is treated as a
dressed Dirac particle where the influence of the coupling to 
the surrounding nucleons via the T-matrix serving as an effective interaction
is expressed by the nucleon self-energy.
The self-energy has a Lorentz
structure with large scalar and four-vector components and
modifies the single-particle spectrum and the spinor wave functions.

The relativistic Brueckner model has been developed in the last decades 
by various authors. At the beginnig of the 80's the relativistic 
problem was considered by the Brooklyn group \cite{23,anastasio83} 
in first order perturbation theory. In the following the 
covariant formalism to a self-consistent 
treatment of the Thompson equation was developed by 
Horowitz and Serot \cite{horse84} 
and is in detail outlined in Ref. \cite{1}. However, in these 
works the nucleon-nucleon interactions was described within the framework 
of the $\sigma\omega$-model. Calculations with realistic  
nucleon-nucleon interactions have been performed by Brockmann and Machleidt 
\cite{brock84,5} and later on by ter Haar and Malfliet \cite{6,7}. 
A more rigorous derivation of the Brueckner approach in
the framework of relativistic Green's function can be found in Ref. \cite{8}. 
In the present work we refer to the formalism described in Refs. 
\cite{1,6}.

The main difference between the relativistic Brueckner 
approach of Refs. \cite{horse84,1,brock84,5,6,7} and  previous works
is the use of medium dependent spinors
and the requirement of self-consistency
for both the spinor wave functions and the single-particle spectrum.  
The Dirac spinors of the nucleon in the medium differ considerably
from the free spinors due to the large scalar and vector self-energies.
These introduce an additional
density dependence into the spinor matrix elements of the one-boson exchange
potentials and thus into the T-matrix. This dynamical effect is absent in 
non-relativistics. 
Another new aspect of the modern Brueckner approach is that 
the inertial frames where the T-matrix is determined 
and where it is used have to be distinguished carefully.
Due to the neglection of the
retardation in the meson propagators the one-boson exchange potentials
and the correspondingly T-matrix serve as quasi-potentials in the 
two-particle center-of-mass (c.m.) frame.
Therefore the relativistic Thompson equation is
solved in this system which can be done by numerical standard techniques,
as, e.g., in the non-relativistic case of Ref. \cite{11}. 
On the other hand single-particle quantities like the 
self-energy of the nucleon are 
calculated in the nuclear matter rest frame, where the single-particle
distribution, the Fermi sea, is naturally defined.
Thus a covariant representation of the T-matrix is required to connect
both frames. 
Even though, e.g., the single particle potential can be 
calculated with some additional approximations avoiding this representation,
as done in Ref. \cite{5}, the covariant technique is unavoidable in order 
to calculate the self-energy components themselves.

In the relativistic Brueckner approach the self-consistent spinors 
are restricted to positive energy states. 
A different approach in  the full Dirac space which is also based on 
one-boson exchange potentials is used in Refs. \cite{23,20}. 
But in a full treatment  one should further include 
renormalisation effects like the vacuum fluctuation contributions 
to the self-energy, as discussed in the context of the relativistic Hartree
approximation in Refs. \cite{3,1}, and solve the full Bethe-Salpeter 
equation self-consistently. However, then the one-boson exchange potentials 
would be inappropriate. But these are still outstanding problems. 
In the standard Brueckner model which we discuss in the present work 
one gets around the problem of renormalization.
The masses and coupling constants in the model are physical renormalized
quantities and as a consequence one neglects the effective 
negative-energy states. However, the effective positive-energy spinors  
still contain admixtures of free negative-energy states because
they are a superposition of free positive- and negative-energy spinors.
To Lorentz transform the T-matrix as we have mentioned above
the positive energy on-shell T-matrix is projected onto 
five invariant amplitudes. 
By the restriction to positive energy states 
an ambiguity arises \cite{24} which is discussed in terms of the pseudo-scalar 
and the pseudo-vector representation of the T-matrix in the Dirac space.
From analogy to the one-pion exchange coupling in the nucleon-nucleon 
potential
the pseudo-vector choice is more natural and further
it is known from meson theory analysis \cite{17} that
this choice suppresses the coupling to the negative-energy states. 
However, the difference between both choices also states a measure for 
the uncertainty concerning the determination of the self-energy 
components due to the restriction 
on positive energy spinors. One issue of the present work is to 
estimate this uncertainty.   

Thus, the aim of this work is to determine the self-energy components, 
to estimate their uncertainty due to different covariant representations of 
the T-operator and different nucleon-nucleon potentials and to examine 
the sensitivity of some commonly used observables, i.e., the equation of state
and the optical potential on the self-energies.
We are investigating in detail the density and the momentum dependence of the 
scalar and the vector components and compare our results to those of other
groups \cite{5,1,6,9}. 
Thereby the momentum dependence 
can just be determined  in a first approximation because 
 it is neglected in the self-consistency treatment of the Brueckner model.
Commonly this approximation is justified by the assumption of a weak
momentum dependence, at least inside the Fermi sea, but there 
have been also indications that this assumption depends on the chosen
one-boson exchange potential \cite{24}. 
We show here for the first time the momentum dependence over its
full range, i.e., below and above the Fermi surface, and we find it 
in the case of the Bonn A,B,C  potentials of the order of several
100 MeV, i.e., to be strikingly strong. 
In difference to the most 
publications in the field we focus our discussion to the self-energy 
components themselves.
 
It is meanwhile well known that the nuclear matter saturation properties
are much better reproduced in the relativistic than in the non-relativistic 
Brueckner approach, i.e., the predictions for the saturation point of a 
variety of nucleon-nucleon potentials are located along the {\it Coester band}
which meets -- in the relativistic case -- the empirical area \cite{5}.
The same also holds, e.g., for the nucleon optical potential.
However, as we also find in the present analysis these commonly studied
observables are not really sensitive to the self-energy components
since they mainly depend on a cancellation of the components.
The situation changes in the application of the Brueckner theory
beyond infinite nuclear matter.
When applying relativistic Brueckner results to the description of
finite nuclei, e.g., in the framework of an effective (density dependent)
relativistic mean field theory \cite{25} the self-energies start to play
a decisive role. This fact is, however, most pronounced in the 
investigation of heavy ion collisions.
Different from resting nuclear matter here the scalar and vector 
components are out of balance
which is due to their different Lorentz transformation character
and diverse observables are in particular sensitive to the magnitude
of the fields \cite{22}. Furthermore the momentum space configuration
can be highly anisotropic and rather corresponds to two counterstreaming
currents of nuclear matter which can be described by two Lorentz-elongated
Fermi ellipsoids. Approximate determinations of relativistic Brueckner 
mean fields for such configurations have been performed in \cite{13}
and encouraging applications to heavy ion collisions in \cite{26}.
Thus the density and momentum dependence of the nuclear self-energy 
components as the quantity which most prominant characterizes the
medium effects is a question of particular interest.

This paper is organized in the following way: in the next section
we review the relativistic Brueckner approach. There are still some
technical differences in the {\it standard} Brueckner model between  
Ref. \cite{1} and Refs. \cite{6,7,8}. Here we follow the approach of 
Horowitz and Serot \cite{1} which allows the use of standard 
techniques for the solution of the Thompson equation 
known from earlier works \cite{11,2}. However, we use the modern 
one-boson exchange potentials, treat the isospin channels 
and also the real and imaginary part of the T-matrix separately 
and project on both, the pseudo-scalar and the pseudo-vector 
representation.  
In the following section we discuss our numerical results. 
The calculations with the $\sigma\omega$-potential are compared 
with results from Ref. \cite{1}. 
We further discuss the density and momentum dependence
of the scalar and vector self-energy, the influence of pseudo-scalar
and pseudo-vector choice is analyzed and we compare to the results of
Brockmann and Machleidt \cite{5} and the Groningen group
\cite{6,9}. Further the equation of state and the optical potential
are shown and their sensitivity to the self-energy components is analyzed.
In the final section a summary is given.

\section{Relativistic Brueckner approach}
In the relativistic Brueckner approach the nucleon inside the nuclear medium 
may be viewed as a dressed particle in conseqeunce 
of its two-body interaction with the surrounding nucleons.
This problem is stated as a coupled set of three non-linear integral equations 
\begin{eqnarray}                                             
   T      & = & V + i \int V Q G G T                 ,    \label{eq1}     \\  
   G      & = & G^{\circ} + G^{\circ} \Sigma G       , \label{eq2}     \\
   \Sigma & = & - i \int_F ( Tr [G T ] - G T )  .            \label{eq3}
\end{eqnarray}
The T-matrix is calculated in the ladder approximation of the Bethe-Salpeter 
equation (\ref{eq1}) and the bare nucleon-nucleon interaction is described 
by one-boson exchange potentials $V$. 
The intermediate nucleons are given by the two-body propagator $iGG$
which is, however, usually replaced by an effective propagator \cite{15}, 
i.e., the Blankenbecler-Sugar propagator or the Thompson propagator.
In the present work the Thompson choice is applied.
The Pauli operator $Q$ accounts for the influence of the medium by the 
Pauli principle and projects the intermediate scattering states 
outside the Fermi sea. 
The dressed one-body Green propagator $G$ is calculated via the Dyson 
equation (\ref{eq2}) from the free propagator $G^{\circ}$. 
The influence of the coupling to the surrounding nucleons is expressed by 
the self-energy $\Sigma$ of the nucleon. 
In the Brueckner formalism this self-energy is determined by summing up the 
direct and exchange interactions with all the nucleons inside the Fermi sea 
$F$, see Eq. (\ref{eq3}). Here the T-matrix plays the role of an 
effective medium dependent two-body interaction. In the relativistic 
Brueckner approach one introduces additional approximations for solving the 
equations (\ref{eq1}-\ref{eq3}) which are sketched in the following, 
for further reading see, e.g., Refs. \cite{1,6}.
\subsection{Self-consistent spinor basis}
The most general form of the Lorentz-structure of the self-energy 
compatible with translational invariance, hermiticity, parity conservation,
time-reversal invariance and rotational invariance (in the nuclear matter 
rest frame) is
\begin{equation}                                            \label{eq4}
    \Sigma(k)  = \Sigma_s(k) - \Sigma^{\mu}(k) \gamma_{\mu} 
     =  \Sigma_s(k) - \gamma_0 \Sigma_0(k) 
     + \mbox{\boldmath$\gamma$\unboldmath} \cdot{\bf k}\Sigma_v(k)   
\end{equation}
with a scalar part $\Sigma_s$ and a vector part 
$\Sigma^{\mu}=(\Sigma_{0},{\bf k} \Sigma_v)$ .
Due to the Dyson equation (\ref{eq2}) the full Green function has 
the formal solution
	\begin{equation}                                         \label{eq5}
	G^{-1}  =  G^{\circ -1} - \Sigma(k)           
	 =  \gamma_{\mu} k^{\ast \mu} - M^{\ast}(k) - i {\rm Im}[\Sigma(k)] 
	\end{equation}
where we have introduced the effective mass and the kinetic momentum
	\begin{equation}                                       \label{eq6}
	M^{\ast}(k) = M + {\rm Re}\Sigma_s(k) \, , \qquad 
	k^{\ast\mu} = k^{\mu} + {\rm Re}\Sigma^{\mu}(k) .
	\end{equation}
Re and Im denote real and imaginary part since  (above the Fermi 
surface) the self-energy in general will be complex.
Here we adopt the {\it quasiparticle approximation} \cite{8}, i.e.,
the Im $[\Sigma]$ will be neglected in Eq. (\ref{eq5}). 
This means that the decay width of the dressed nucleon is set equal to zero, 
resulting in an infinite lifetime of this 'quasiparticle'.
In the relativistic Brueckner approach the full Green function $G$ 
is replaced by its positive energy part.
Thus the self-energy in Eq. (\ref{eq3}) is determined by the 
part proportional to the Fermi sea $F$, 
the so-called Dirac Green function $G_D$ \cite{3,1}
\begin{equation}                                          \label{eq7}
G_D(k) = [\gamma_{\mu} k^{\ast\mu} + M^{\ast}(k)]2\pi i 
\delta(k^{\ast 2} - M^{\ast 2}(k))\Theta(k^{\ast 0}) \Theta(k_F-|{\bf k}|) 
\end{equation}
which eliminates the divergent contributions from the negative energy sea.
However, in the form of Eq. (\ref{eq7}) $G_D$ is restricted to the 
description of nuclear matter at rest. In order to achieve general convenience
the Fermi sphere $\Theta(k_F-|{\bf k}|)$ has to be 
replaced by an Fermi ellipsoid $\Theta(E_F - k^{\ast\mu} u_{\mu})$ with the 
Fermi energy $E_F = \sqrt{ (1+\Sigma_v)^2 {k_F}^2 + M^{\ast 2}}$ and 
the boost 4-velocity $u^{\mu} = (\gamma, {\bf u}\gamma)$ \cite{13}.
In Eq. (\ref{eq7}) the momenta of the quasiparticles are put on mass-shell,
which is expressed by the $\delta$-distribution and thus 
$k^{\ast 0}=E^{\ast}({\bf k})=\sqrt{ {\bf k}^{\ast 2} + M^{\ast 2}}$.  
Hence the self-energy $\Sigma(k)=\Sigma(|{\bf k}|,k_F)$ depends only on the 
three-momentum $|{\bf k}|$ and, of course, on the respective density or $k_F$.

Furthermore, in the relativistic Brueckner model one applies a mean field 
approximation to $\Sigma$, Eq. (\ref{eq4}), which allows a simple formulation 
of the self-consistency problem. The explicit momentum dependence of the 
self-energy which enters via the term ${\bf k} \Sigma_v$
can be dealt with by introducing the reduced kinetic
momentum $\tilde k^{\ast\mu} = k^{\ast\mu} / (1+\Sigma_v)$ and the reduced 
effective mass $\tilde M^{\ast} = M^{\ast} / (1+\Sigma_v)$.
Further one neglects the momentum dependence of 
the effective mass ($M^{\ast}$ or $\tilde M^{\ast}$). 
Thus, the nucleons fulfill a quasi-free Dirac equation 
\begin{equation}                                     \label{eq8}
	\left[ \gamma_{\mu} \tilde k^{\ast\mu} - \tilde M^{\ast} \right] 
	u_{\lambda}(k) = 0 
\end{equation}
and using the normalization of Ref. \cite{14} 
the self-consistent positive-energy spinors of helicity $\lambda$ 
are defined as
\begin{equation}                    \label{eq9}
  u_{\lambda}(k) = \sqrt{\frac{\tilde E^{\ast}(k) + \tilde M^{\ast}}
                            { 2 \tilde M^{\ast}}                  }
  \left( \begin{array}{c}  1  \\
      \frac{2 \lambda |{\bf k}| }{\tilde E^{\ast}(k) + \tilde M^{\ast}} 
         \end{array} 
  \right) \chi_{\lambda}
\end{equation}
with $\chi_{\lambda}$ being a Pauli spinor. The Dirac spinors depend on 
the effective mass and thus on the nuclear density. 
The matrix-elements of the 
T-matrix and the one-boson exchange potentials $V$, Eq. (\ref{eq1}), are 
calculated with these spinors.
In contrast to non-relativistic Brueckner theory this fact introduces 
an additional density dependence into the interaction 
which is one major reason
for the great success of the relativistic treatment.
This density dependence is mediated by the 
additional parameter $ \tilde M^{\ast}$ which is fixed at a reference
point, usually at $|{\bf k}|=k_F$.
$ \tilde M^{\ast}$ is obtained as the solution of the non-linear equation
({\it reference spectrum approximation}) 
	\begin{equation}                                    \label{eq10}
	\tilde M^{\ast} = M + \Sigma_s(k_F,\tilde M^{\ast}) 
	- \tilde M^{\ast} \Sigma_v(k_F,\tilde M^{\ast}) 
	\end{equation}
which follows from the definition given above.
Self-consistency is achieved by determining for a start value of 
$\tilde M^{\ast}$ the medium-dependent spinors of Eq. (\ref{eq9}), 
next the T-matrix from Eq. (\ref{eq1}) 
and finally the self-energy and the new $\tilde M^{\ast}$\footnote{From now
on we omit the tilde because in the following we normally deal with 
$\tilde M^{\ast},\tilde k^{\ast\mu}$.}.
This procedure is repeated until convergence is reached.
\subsection{Covariant T-matrix amplitudes}
The T-matrix, in Eq. (\ref{eq1}), is most easily determined in the two-particle
center-of-mass (c.m.) system while the self-energy, Eq. (\ref{eq3}), 
is calculated
in the nuclear matter rest frame. Therefore we will determine 
in a second step covariant amplitudes of the T-matrix.
The Thompson propagator (and similar the 
Blankenbecler sugar propagator) projects the intermediate nucleons 
onto positive energy states and restricts the exchanged energy
transfer by $\delta(k^0)$ to zero. Thus  
equation (\ref{eq1}) is reduced to a three dimensional integral equation of
the Lippmann-Schwinger type, the so called Thompson equation \cite{15,6}
\begin{equation}                   \label{eq11}
T({\bf p},{\bf q},x)|_{\rm c.m.} = V({\bf p},{\bf q}) 
+ \int \frac{d^3 {\bf k}}{(2\pi)^3}  V({\bf p},{\bf k}) 
\frac{M^{\ast 2}}{ E^{\ast 2}({\bf k}) }
\frac{Q({\bf k},x)}{ 2E^{\ast}({\bf q}) - 2E^{\ast}({\bf k}) 
+ i\epsilon } T({\bf k},{\bf q},x)      .
\end{equation}
Here the Thompson equation is given in the two-nucleon c.m.-frame where
${\bf p}=({\bf p}_1-{\bf p}_2)/2
=({\bf p}^{\ast}_1-{\bf p}^{\ast}_2)/2$ 
is the relative momentum of the final states and similar ${\bf q,k}$ 
are the relative momenta of the initial and intermediate states, respectively.
The Thompson propagator and similar the Blankenbecler-Sugar propagator 
imply that the timelike component of the momentum transfer in $V$ and $T$ 
is set equal to zero which is a natural constraint in the c.m.-frame,
however, not a covariant one.
Hence we solve the Thompson equation in the c.m.-frame
in contrast to, e.g., the approach of Ref. \cite{5}. 
The starting energy in Eq. (\ref{eq11}) is already fixed: 
$ \sqrt{s^{\ast}} = 2E^{\ast}({\bf q})$.
The Pauli operator $Q$ explicitely 
depends on the chosen frame, i.e., on  the 
boost 3-velocity ${\bf u}$ into the c.m.-frame.
With $x$ we denote the set of parameters  
$x=\{k_F,M^{\ast},|{\bf u}|\}$ on which the T-matrix actually depends.

We solve the Thompson equation (\ref{eq11}) for the on-shell T-matrix
$(|{\bf p}|=|{\bf q}|)$ in the c.m.-system and thereby apply standard 
techniques 
which are described in detail by Erkelenz \cite{2}. 
We construct the positive-energy helicity T-matrix elements from the 
$|JMLS>$-scheme. In the on-shell case 
only five of the sixteen helicity matrix elements are independent
which follows from general symmetries, see Ref. \cite{2}.
After a partial wave 
projection onto the $|JMLS>$-states the integral reduces to a one-dimensional 
integral over the relative momentum $|{\bf k}|$ and Eq. (\ref{eq11}) decouples
into three subsystems of integral equations for the uncoupled spin singlet,
the uncoupled spin triplet and the coupled triplet states.
The Pauli operator $Q$ is replaced by an angle averaged Pauli operator
$\overline Q$ \cite{1}. 
Since the Fermi sphere is deformed to a Fermi ellipsoid in the two-nucleon 
c.m.-frame $\overline Q$ is evaluated for such a Fermi-ellipsoid.
\begin{equation}                     \label{eq11a}
{\overline Q}(|{\bf k}|,x) 
= \left\{
       \begin{array}{lcl}
        0    &     & |{\bf k}| < k_{-} \\
        (\gamma E^{\ast}({\bf k}) - E_F) / u \gamma |{\bf k}|  
             & for &  k_{-} \le |{\bf k}| \le k_{+}  \\
        1    &     &  |{\bf k}| > k_{+}
       \end{array}
  \right. 
\end{equation}
with $k_{-}=\sqrt{k_F^2 - u^2 E_F^2}, \; k_{+}=\gamma(u E_F + k_F)$ 
and $u=|{\bf u}|$.
We are solving the integral equations by the
matrix inversion techniques of Haftel and Tabakin \cite{11}. 
Real and imaginary parts of the T-matrix are calculated separately 
by the principal-value treatment given in  Ref. \cite{10}. 
Due to the antisymmetry of the two-fermion states we can restore the total
isospin I of the two-nucleon system (I=0,1) with the help of the 
selection rule: 
$(-)^{L+S+{\rm I}}=-1$.
From the five independent on-shell amplitudes in the $|JMLS>$-representation
the five independent partial wave amplitudes in the helicity representation
(for I=0,1 and real and imaginary part separately) are obtained
by inversion of Eq. (3.32) and then of Eq. (3.28) of Ref. \cite{2}.
The summation over the total angular momentum $J$ (we have taken $J\leq 6$
and for comparison  $J\leq 12$) yields the full helicity matrix element
\begin{equation}                       \label{eq12}
\sum_J \left[ \frac{2J+1}{4\pi} \right]
d^J_{\lambda \lambda^{\prime} }(\theta)
<|{\bf p}|\lambda^{\prime}_1 \lambda^{\prime}_2 |T^{J,{\rm I}}(x)|
\, |{\bf q}| \lambda_1 \lambda_2>
= <{\bf p} \lambda^{\prime}_1 \lambda^{\prime}_2 {\rm I} {\rm I}_3 | T(x)
| {\bf q} \lambda_1 \lambda_2 {\rm I} {\rm I}_3 > .
\end{equation}
Here $\theta$ is the scattering angle between ${\bf q}$ and ${\bf p}$ and 
$\lambda = \lambda_1 - \lambda_2, \lambda' = \lambda_1' - \lambda_2'$.
The reduced rotation matrices $d^J_{\lambda \lambda^{\prime} }(\theta)$ 
are those defined by Rose \cite{16}.
The matrix element (\ref{eq12}) is actually independent of the third
component of the isospin ${\rm I}_3$.
The next step is the projection of the five independent helicity amplitudes
onto five covariant amplitudes. Therefore the T-operator is expanded 
into the basis matrices in the Dirac spinor space 
\begin{equation}                    \label{eq13}
T^{\rm I} = 
T^{\rm S,I} 1_{(1)} 1_{(2)} 
+ T^{\rm V,I} \gamma^{\mu}_{(1)} \gamma_{(2)\mu}          
+ T^{\rm T,I} \sigma^{\mu\nu}_{(1)}\sigma_{(2)\mu\nu}
+ T^{\rm P,I} 
 \left( \frac{{\not \rm K}\gamma_5}{2M^{\ast}} \right)_{(1)} \!
 \left( \frac{{\not \rm K}\gamma_5}{2M^{\ast}} \right)_{(2)} 
+ T^{\rm A,I} (\gamma_5 \gamma^{\mu})_{(1)} (\gamma_5 \gamma_{\mu})_{(2)} 
\end{equation}
The subscripts $(1),(2)$ denote the particle on which the matrix acts.
This expansion, however, is not unique since we have specified 
the Dirac matrix structure of the T operator 
by its action only on positive energy states.
The expansion of Eq. (\ref{eq13}) is given with  
 $T^{\rm P,I}$ a pseudo-vector interaction
${\not \rm K}\gamma_5$. It is defined as 
${\not \rm K}=\gamma^{\mu}{\rm K}_{\mu}$ with
the transferred momentum at the vertex ${\rm K}^{\mu}=p^{\mu}_f-p^{\mu}_i
=p^{\ast\mu}_f-p^{\ast\mu}_i$ where $f,i$ indicates final and initial 
momentum. We will, however, also adopt a pseudo-scalar interaction 
where the fourth term in Eq. (\ref{eq13}) is replaced by
$T^{\rm P,I} \gamma_5^{(1)}\gamma_5^{(2)}$ as it was also done 
by Horowitz and Serot \cite{1}.
The pseudo-vector choice is, e.g., made by the Groningen group. 
This choice is supported by the argument that 
it agrees with the one-pion exchange coupling
in the nucleon-nucleon potential
and furthermore suppresses the coupling to the negative energy states 
which are neglected in the Brueckner approach \cite{6,17}. In the present work 
we apply both variants since we want to examine the influence 
of the different choices on the self-energy.

Next we take the helicity matrix elements of Eq. (\ref{eq13}). On the r.h.s.
the helicity states are acting directly on the matrix operators which are now
abbreviated by $\kappa_m^{(i)}$ for i=1,2 and $m\varepsilon\{S,V,T,P,A\}$
\begin{equation}                    \label{eq14}
<{\bf p} \lambda^{\prime}_1 \lambda^{\prime}_2 {\rm I} {\rm I}_3 | T(x)
| {\bf q} \lambda_1 \lambda_2 {\rm I} {\rm I}_3 > = 
\sum_m  <{\bf p} \lambda^{\prime}_1 \lambda^{\prime}_2| 
\kappa_m^{(1)} \kappa_m^{(2)} | {\bf q} \lambda_1 \lambda_2>
T^{m,\rm I}(|{\bf p}|, \theta,x)  .
\end{equation}
The helicity matrix elements on the r.h.s. are given explicitely, e.g., in 
Eqs. (2.11-2.17) of Ref. \cite{2}. For on-shell scattering  
between positive energy states the matrix elements of the 
pseudo-scalar (PS) and the pseudo-vector (PV) matrix operators are identical 
\cite{17}.
This vertex equivalance follows immediately from the Dirac Eq. (\ref{eq8}):
${\overline u}(p)\left( \frac{{\not p}^{\ast}-{\not q}^{\ast}}{2M^{\ast}}  
\right) \gamma_5 u(q) = {\overline u}(p)\gamma_5 u(q)$.
Thus the covariant amplitudes $T^{\rm PS}=T^{\rm PV}\equiv T^{\rm P} $ 
are identical as well. 
The difference 
between both representations of the T-matrix in Dirac space enters
only at the one-body level when calculating the self-energy \cite{19}. 
For fixed parameters $(x,|{\bf p}|=|{\bf q}|,\theta)$ equation (\ref{eq14})
is a matrix relation between the five independent 
helicity amplitudes 
$T^{\rm I}_i$ (five of the sixteen amplitudes numbered with 
$i=\{\lambda^{\prime}_1 \lambda^{\prime}_2;\lambda_1 \lambda_2 \}=1,..,5$) 
on the l.h.s. and the five unknown covariant amplitudes $T^{m,\rm I}$
\begin{equation}                    \label{eq15}
T^{\rm I}_i = \frac{1}{M^{\ast 2}}\sum_m C_{i m}T^{m,\rm I}  .
\end{equation}
This corresponds to Eq. (3.23) of Ref. \cite{1} with respect to our different
normalization of the spinor basis, Eq. (\ref{eq9}).
The covariant amplitudes are determined by a matrix inversion 
of equation (\ref{eq15}).
This, however, only has to be done for two angles since
for on-shell scattering in the two-particle c.m.-frame only two
scattering angles appear, i.e., $\theta=0$ for the direct and $\theta=\pi$ for 
the exchange interaction. 
However, now the matrix becomes singular because two of the 
five helicity amplitudes vanish. Therefore it is useful to extract
the leading angular dependence of the matrix $C_{i m}$ as well as 
of the helicity amplitudes $T^{\rm I}_i$  from the rotation 
matrices $d^J_{\lambda \lambda^{\prime} }(\theta=0,\pi)$ in Eq. (\ref{eq12}). 
Thus, the limit $\theta \rightarrow 0,\pi$ can be performed analytically. 
This method has been developed in appendix C of Ref. \cite{1} and formula (\ref{eq15})
can be found there for the limiting angles $0,\pi$ (Eqs. (C.10,11)). 
Proceeding this way we determine the real and the imaginary part of the 
five covariant amplitudes for the direct and exchange contribution in both, 
the isospin singlet and
triplet channels of the T-matrix $T^{m,\rm I=0,1}(\theta=0,\pi)$. 
\subsection{Self-energy components}
Now we are able to calculate the self-energy. Inserting the Green function, 
Eq. (\ref{eq7}), and the Dirac representation of the T-matrix, 
Eq. (\ref{eq13}), into 
the definition of the self-energy, Eq. (\ref{eq3}), yields 
\begin{eqnarray}                                     \nonumber
 \Sigma_{\alpha\beta} =  
 \int \frac{d^3{\bf q}}{(2\pi)^3} 
\frac{\Theta(k_F-|{\bf q}|)}{4E^{\ast}({\bf q})}
 \left\{ M^{\ast} 1_{\alpha\beta}
\left[  4 T^{\rm S}_D - T^{\rm S}_X -4 T^{\rm V}_X 
               - 12 T^{\rm T}_X + 4 T^{\rm A}_X  
               - \frac{(k^{\ast\mu} - q^{\ast\mu})^2}{4M^{\ast 2}}
               T^{\rm P}_X\right]    \right.              \\ \label{eq16a}    
 \left.  + {\not q}^{\ast}_{\alpha\beta} 
\left[ 4 T^{\rm V}_D - T^{\rm S}_X +2 T^{\rm V}_X 
              + 2 T^{\rm A}_X  
 - \frac{(k^{\ast\mu} - q^{\ast\mu})^2}{4M^{\ast 2}}
                         T^{\rm P}_X\right]             
  - ({\not q}^{\ast} - {\not k}^{\ast})_{\alpha\beta}
       \frac{2 q^{\ast}_{\mu} (k^{\ast\mu} - q^{\ast\mu})}{4M^{\ast 2}}
                         T^{\rm P}_X   \right\} 
\end{eqnarray}
where the $\alpha,\beta$ denote both, Dirac (spin) and 
isospin indices. The direct and exchange parts of 
the T-matrix denoted by $D$ and $X$
are calculated from isospin sums of the isospin projection 
operators which yields $T^m_D =T^{m {\rm I}=0}_D+3T^{m {\rm I}=1}_D $
and $T^m_X = -T^{m {\rm I}=0}_X+3T^{m {\rm I}=1}_X $ .
The self-energy as a one-body operator in spinor space posseses components 
proportional to the unity matrix 1 and to $\gamma^{\mu}$ as 
postulated in equation (\ref{eq4}). From this relation 
the self-energy components $\Sigma_{s,0,v}$ follow 
by taking traces
\begin{eqnarray}                                     \label{eq16}
\Sigma_s(|{\bf k}|;M^{\ast},k_F) & = & \quad \! \frac{1}{8} Tr[\Sigma]
\qquad =\int \frac{d^3{\bf q}}{(2\pi)^3} \Theta(k_F-|{\bf q}|)
\frac{M^{\ast}}{E^{\ast}({\bf q})} T_s({\bf k,q};x)     , \\  \label{eq17}
\Sigma_{0}(|{\bf k}|;M^{\ast},k_F) & = & 
- \frac{1}{8} Tr[\gamma^{0} \Sigma]         
\quad =\int \frac{d^3{\bf q}}{(2\pi)^3} \Theta(k_F-|{\bf q}|)
T_{(0)}({\bf k,q} ;x)                                   ,   \\   \label{eq18}
\Sigma_{v}(|{\bf k}|;M^{\ast},k_F) & = & 
- \frac{1}{8} Tr[\mbox{\boldmath$\gamma$\unboldmath} \cdot{\bf k} \Sigma]
=\int \frac{d^3{\bf q}}{(2\pi)^3} \Theta(k_F-|{\bf q}|)
\frac{{\bf q}^{\ast}\cdot{\bf k}}{|{\bf k}|^2 E^{\ast}({\bf q})} 
T_{v}({\bf k,q};x) 
\end{eqnarray}
where we have used the abbreviations
\begin{eqnarray}                                                \label{eq19}
T_s({\bf k,q} ;x) & = &
\frac{1}{4}\left[ 4 T^{\rm S}_D - T^{\rm S}_X -4 T^{\rm V}_X 
               - 12 T^{\rm T}_X + 4 T^{\rm A}_X  
               + \frac{ k^{\ast}_{\mu} q^{\ast\mu} - M^{\ast 2}}{2M^{\ast 2}}
                                           T^{\rm P}_X\right]  ,\\ \label{eq20}
T_{(0)}({\bf k,q} ;x) & = &
\frac{1}{4}\left[- 4 T^{\rm V}_D + T^{\rm S}_X -2 T^{\rm V}_X 
               - 2 T^{\rm A}_X  
 + \frac{E^{\ast}({\bf k})}{E^{\ast}({\bf q})}  
   \frac{M^{\ast 2} - k^{\ast}_{\mu} q^{\ast\mu}}{2M^{\ast 2}}
                                           T^{\rm P}_X\right]  ,\\ \label{eq21}
T_{v}({\bf k,q} ;x) & = &
\frac{1}{4}\left[- 4 T^{\rm V}_D + T^{\rm S}_X -2 T^{\rm V}_X 
               - 2 T^{\rm A}_X  
 + \frac{|{\bf k}|}{q_z}  
   \frac{M^{\ast 2} - k^{\ast}_{\mu} q^{\ast\mu}}{2M^{\ast 2}}
                                           T^{\rm P}_X\right]  .
\end{eqnarray}
In the pseudo-scalar case \cite{1} all the cofactors of $T^{\rm P}_X$
are replaced just by -1 which yields $T_{(0)}=T_{v}$. The self-energy of a 
nucleon is calculated in the nuclear matter rest frame by integrating 
the effective two-body interaction over the Fermi sea. 
The 3-momenta ${\bf k,q}$ of
a nucleon pair are given in this frame and we define
${\bf P}^{\ast} = {\bf k} + {\bf q}= {\bf k}^{\ast} + {\bf q}^{\ast}$. 
On the other hand the covariant amplitudes are determined in the two-nucleon 
c.m. system and thus depend on the respective 
quantities in that frame
\begin{eqnarray}                                                \label{eq22}
T^m_D({\bf k}{\bf q},{\bf k}{\bf q};x) & = &
\quad T^{m {\rm I}=0}(|{\bf p}_{\rm c.m.}|, \theta=0,x) 
+ 3 T^{m {\rm I}=1}(|{\bf p}_{\rm c.m.}|, \theta=0,x)      , \\ \label{eq23}
T^m_X({\bf k}{\bf q},{\bf q}{\bf k};x) & = &
- T^{m {\rm I}=0}(|{\bf p}_{\rm c.m.}|, \theta=\pi,x) 
               + 3 T^{m {\rm I}=1}(|{\bf p}_{\rm c.m.}|, \theta=\pi,x)   .
\end{eqnarray}
From the invariant mass
$s^{\ast} = ( E^{\ast}({\bf k}) + E^{\ast}({\bf q}) )^2 - {\bf P}^{\ast 2}$,
 the relative momentum in the c.m.-frame follows by
$|{\bf p}_{\rm c.m.}| = \sqrt{ s^{\ast}/4 - M^{\ast 2} }$. 
As already mentioned above, $x$ abbreviates the set of parameters 
$x=\{k_F,M^{\ast},|{\bf u}| \}$, 
where the boost 3-velocity (from the nuclear matter into the c.m.-frame) 
is now given by
${\bf u} = {\bf P}^{\ast} / \sqrt{ s^{\ast} + {\bf P}^{\ast 2} }$.
In the two-particle c.m.-frame there exist only two possible scattering 
angles, 
i.e., $\theta=0$ for the direct and $\theta=\pi$ for the exchange amplitude.
Numerically we solve the Thompson equation (\ref{eq11})
and determine the covariant amplitudes $T^{m,\rm I}$ for a suitable
two-dimensional range of the quantities ${\bf p}_{\rm c.m.},|{\bf u}|$. 
Using the azimuthal
symmetry we calculate the self-energy, Eqs. (\ref{eq16}-\ref{eq18}), 
by a two-dimensional integral over the Fermi sphere. 
Therefore for each nucleon pair we have to interpolate 
the covariant amplitudes at the respective values of ${\bf p}_{\rm c.m.}$
and $ |{\bf u}|$. 
The resulting self-energy, taken at the
Fermi surface $|{\bf k}|=k_F$, yields 
the new value of the effective mass $M^{\ast}$, Eq. (\ref{eq10}).
This value serves as the input for the next iteration. 
This procedure is repeated until a convergence of the effective 
mass is achieved.

\section{Nuclear matter results}
For the solution of the Thompson equation (\ref{eq11}) we apply 
the Bonn A,B,C potentials of Refs. \cite{4,5} as 
the nucleon-nucleon interaction $V$. 
The potentials are calculated 
with the OBNNS code of R. Machleidt \cite{machleidt93}. 
The Bonn potentials are based on the exchange of 
the six nonstrange bosons with masses below 1 GeV 
$(\pi, \eta, \rho, \omega, \delta, \sigma)$. For the pion a derivative
pseudovector coupling is applied. The three parametrizations A,B and C differ 
essential in the $\pi$NN form-factor and as a consequence in 
the strength of the nuclear tensor force. 
Actually we find  only minor differences of a few
MeV in the resulting self-energies, see below, 
and thus we present mainly the Bonn C results. 
For comparison with the results of Horowitz and Serot 
we also employ the simple $\sigma\omega$-meson exchange  
of Ref. \cite{1} with $g_{\sigma}^2=91.64, \; g_{\omega}^2=136.2$ 
and use the form-factor given therein.
\subsection{self-energy components}
\subsubsection{density dependence}
First we want to demonstrate the reliability of our approach 
by comparison with the results of Ref. \cite{1}.
In Fig.\,1 we show the self-consistent effective mass $M^{\ast}$, 
Eq. (\ref{eq10}), as a function of the Fermi momentum $k_F$ which is a measure
for the density $\varrho = 2/(3\pi^2) k_F^3$. Our results with the
$\sigma\omega$-meson exchange potential and the pseudo-scalar decomposition
of the T-matrix reproduce quite well the values
taken from Ref. \cite{1}. The slight deviation for densities above 
saturation density (here $k_F=1.42\,fm^{-1})$ is partly due to the different
effective two-nucleon propagators. As mentioned above, we are using 
the Thompson propagator, Eq. (\ref{eq11}), while in Ref. \cite{1} the 
Blankenbecler-Sugar propagator is applied. Furthermore, in Ref. \cite{1} 
an additional approximation has been made. As discussed above, we are solving 
the Thompson equation for a two-dimensional grid of the relative
momentum $|{\bf p}|$ and the c.m.-momentum $|{\bf P}|$ respectively 
the boost-velocity $|{\bf u}|$ and
use this T-matrix for the determination of $\Sigma$. In Ref. \cite{1} the
$|{\bf P}|$-dependence is replaced by an "average $|{\bf P}|$ approximation" 
for each $|{\bf p}|$ which causes deviations mainly at high density systems.
For the Bonn C potential the effective mass is significantly different
from that of the $\sigma\omega$-model. 
Here we applied the pseudo-vector choice.

In Fig.\,2 we show the density dependence of the self-energy, i.e.,
the scalar part $\Sigma_s$ as well as the time-like $-\Sigma_0$ and 
the space-like component $-k_F\Sigma_v$ of the vector self-energy. 
The self-energies are determined for the three Bonn potentials A,B,C
and again we the pseudo-vector choice is applied.
For the sake of a better comparison
the sign of the vector  self-energies is inversed and 
the dimensionless space-like component $\Sigma_v$ is multiplied by $k_F$. 
The scalar and also the vector
self-energy component $\Sigma_s,-\Sigma_0$ differ only slightly for 
the three Bonn potentials. The difference amounts about
5 MeV at saturation density (here $k_F = 1.35 fm^{-1}$) 
and is still less than 10 MeV at $k_F = 1.90 fm^{-1}$. In absolute magnitude
Bonn C yields the weakest and Bonn A the strongest self-energies.
The space-like vector self-energy $-k_F\Sigma_v$ is equal within
line-width for the three Bonn potentials and is small in comparison to
$\Sigma_s,\Sigma_0$, e.g., about 7\% of $\Sigma_0$ at saturation density.
For further analysis we therefore concentrate on  the large components
$\Sigma_s,\Sigma_0$ calculated with the Bonn C potential.
\subsubsection{pseudo-scalar versus pseudo-vector choice}
The two different representations of the T-matrix, i.e., the pseudo-scalar
and the pseudo-vector case, represent an inherent ambiguity.
This is, however, a common feature of all
relativistic Brueckner calculations which specify the Dirac matrix 
structure of the T-matrix by using only positive energy T-matrix elements.  
In order to clarify the influence of this ambiguity in Figs.\,3. and 4.  
we show the self-energy components $\Sigma_{s,0}$ calculated exemplaryly 
with the Bonn C potential for both choices. 
In the pseudo-scalar case, see Fig.\,3, the 
self-energies are larger by about 40-80 MeV (decreasing
with increasing density) than in
the pseudo-vector case (Fig.\,4). 
This difference may be interpreted as
 a measure of the general uncertainty of the self-energy components 
arising from any particular 
projection of the T-operator on positive energy matrix elements only.
This uncertainty is larger at low densities.
To get more insight into the origin of this difference we decompose the 
self-energy in the two contributions stemming from the direct and 
the exchange T-amplitude respectively.
E.g., in the pseudo-scalar case the direct part of the self-energy $\Sigma_s$
is determined by the integral over $T^{\rm S}_D$ 
while the exchange part is given by the integral over the sum
of all exchange-amplitudes 
$- 1/4 T^{\rm S}_X - T^{\rm V}_X - 3 T^{\rm T}_X +  T^{\rm A}_X 
- 1/4 T^{\rm P}_X$,
see Eq. (\ref{eq19}) and the subsequent comment. 
We find that in the pseudo-scalar case direct and exchange parts of
$\Sigma_{s,0}$ are identical within less than 0.5 MeV, see Fig.\,3. 
However, in the pseudo-vector case the factor accompanying 
the $T^{\rm P}_X$ amplitude in Eqs. (\ref{eq19},\ref{eq20}) reduces the 
exchange part by roughly 50\%. 
For comparison in the $\sigma\omega$-model, which lacks of pion exchange, 
the $T^{\rm P}_X$ 
amplitude is generally smaller and the effect, i.e., 
the reduction of the exchange part, amounts to roughly 25\%.
Of course also
the direct part is different from that of the pseudo-scalar case
since we consider the self-consistently determined fields.
We conclude that the pseudo-vector choice in general reduces the total 
self-energy which is due to a considerable reduction of the exchange
contributions  while the direct contributions are even enhanced.

In Fig.\,5 we compare our results for $\Sigma_s,-\Sigma_0$  
obtained with the pseudo-vector choice and the Bonn C potential
with the results of two other groups, i.e., the Groningen group
\cite{6} and the approach of Brockmann and Machleidt \cite{5}. 
In Ref. \cite{5} the relativistic Thompson equation
is solved by standard methods \cite{11} directly in the nuclear 
matter rest frame instead of the two-particle c.m.-frame. 
Therefore the Thompson Eq. (\ref{eq11})
is Galilei transformed to the rest frame with the c.m.-momentum ${\bf P}/2$.
The relativistic single particle potential $U$ in the nuclear matter 
rest frame can be evaluated from this T-matrix by
\begin{equation}
U(|{\bf k}|) = \frac{M^{\ast}}{E^{\ast}({\bf k})} 
\, {\rm Re} <k| \Sigma |k>
= \sum_{q \varepsilon F} \frac{M^{\ast 2}}{E^{\ast}({\bf k})E^{\ast}({\bf q})}
\, {\rm Re} <kq|T|kq - qk>             \label{eq24}
\end{equation} 
with the definition of $\Sigma(|{\bf k}|)$ as in Eq. (\ref{eq3}). The 
advantage of this method is that one is not forced to Lorentz transform 
the matrix elements between the
nuclear matter rest frame and the two-particle c.m.-frame and thus a
projection onto covariant amplitudes is obsolete which simplifies the task
considerably. However, now the self-energy components 
$\Sigma_s,-\Sigma_0$ cannot be calculated directly  as in the projection 
method of Eqs. (\ref{eq16}-\ref{eq18}). Instead,
one has to extract them from the single particle potential $U$. 
With  the form of $\Sigma$ as in Eq. (\ref{eq4}), however, thereby 
neglecting $\Sigma_v$ one finds 
\begin{equation}                         \label{eq25}
U(|{\bf k}|) = \frac{M^{\ast}}{E^{\ast}({\bf k})}\Sigma_s
- \Sigma_0  ,
\end{equation} 
i.e., the single particle potential results from a cancellation of the
huge scalar and vector components.
Using a mean field ansatz for $\Sigma_{s,0}$, i.e., neglecting their
momentum dependence, one is able to determine them by a fit procedure 
to $U(|{\bf k}|)$. 
In Ref. \cite{5} the self-energy components are constructed
by this potential-fit method.
In Fig.\,5  we compare these results with our calculations 
both for the Bonn C potential. In the density range considered the fitted
self-energies are generally smaller in absolute values by 50 up to more 
than 150 MeV, e.g., at saturation density they are reduced by about 75 MeV. 
In spite of this discrepancy both 
calculations have several features in common. E.g., the single particle
potential, defined in Eq. (\ref{eq24}), is almost identical 
 and as we will see later on the same holds for the equation of state. 
Concerning  these two quantities
the main difference between both treatments lies in the reference frame 
where the T-matrix is actually calculated while the projection method 
on covariant amplitudes is not essential. 
At moderate densities the T-matrix
depends only weakly on the c.m.-momentum ${\bf P}$ and respectively 
on the chosen frame.
Thus the numerical results show only minor differences. Another feature
in common is that also the potential-fit method yields self-energies 
for the three types of Bonn potentials which are equal within a few MeV 
\cite{5}.   

In Fig.\,5 we further compare to the results of the Groningen group 
\cite{6}. They also perform
a full relativistic Brueckner calculation solving the Thompson 
equation in the two-particle c.m.-system and using a projection of the
T-matrix onto a pseudo-vector representation. The difference to our method
is on the one hand the application of a different 
one-boson exchange potential including different form-factors, 
the so called Groningen potential.  
On the other hand as - a more technical point - in Ref. \cite{6} the 
Thompson equation is solved in full momentum-spin space by the use of 
Pad\'e approximants, for further details see, e.g., Ref. \cite{8}. 
For densities above saturation density the self-energies 
calculated with the Groningen potential are
roughly the same as the ones obtained by us, however, 
in the low density range they are much smaller in magnitude.  
\subsubsection{momentum dependence}
Up to now we have discussed the self-energy of a nucleon in nuclear matter 
obtained at the Fermi surface ($|{\bf k}|=k_F$), 
i.e., the density dependence of the self-energy. 
In the relativistic Brueckner theory the self-energy at the Fermi surface
and the T-matrix are  determined self-consistently, as described above.
The dressed nucleon propagators respectively the self-consistent spinor basis,
Eq. (\ref{eq9}), entering into the calculation of the T-matrix are constructed
with the effective mass, Eq. (\ref{eq10}). $M^{\ast}$ itself depends on 
the density dependent self-energies  $\Sigma_{s,0,v}(k_F)$. Thus only 
the effective mass acts as a self-consistency or iteration parameter.
In a next step the self-energies above and below the Fermi surface
can be evaluated from Eqs. (\ref{eq16}-\ref{eq18}).
This has to be considered just as the first iteration step in the
determination of  self-consistent
momentum dependent functions $\Sigma_{s,0,v}(|{\bf k}|)$
and stands in contrast to the non-relativistic Brueckner treatment
where self-consistency is required for the full positive 
single particle energy spectrum. 
However, these approximations are not based on the same footing. 
In the relativistic treatment we have an intrinsic 
momentum dependence due to the Dirac structure of the nucleon which 
introduces a momentum dependence in 
the single particle potential $U$ even for constant
values of $\Sigma_{s,0,v}$, see Eq. (\ref{eq25}).
A further argument often given in connection with the reference spectrum 
approximation \cite{1,6} is that the 
momentum dependence of the self-energies in the relativistic
Brueckner approach is found to be rather soft. 
However, in the present work we demonstrate that this is in general 
not the case but strongly depends on the special form of the 
one-boson exchange potential.
In Fig.\,6 we show the momentum dependence of the nucleon self-energies 
$\Sigma_{s,0}$ at saturation density 
calculated with the $\sigma\omega$-exchange potential. 
Our results nicely agree
with the results of Ref. \cite{1} within 10 MeV which might be due to 
different technical details already mentioned in the discussion of
Fig.\,1. We find that $\Sigma_s$ decreases in the range from 
$|{\bf k}|=0$ to $k_F$ by 14 MeV and $\Sigma_{0}$ by 24 MeV, i.e., 
below the Fermi surface the self-energies are constant within a few percent. 

In Fig.\,7 we show the momentum dependence of the self-energies for
calculations with the Bonn C potential (pseudo-vector choice) and the
$\sigma\omega$-potential (pseudo-scalar choice), both at $k_F=0.265$ GeV.
More strictly this is the real part of the self-energy since 
for momenta greater than the Fermi momentum the self-energy is complex.
The results obtained with the Bonn C potential show a strong momentum
dependence: over the entire momentum range up to 0.8 GeV 
corresponding to a single particle energy of around 300 MeV 
the self-energies decrease by nearly 60\% with respect to the central value. 
The most pronounced change occurs in the region around the
Fermi surface while deep inside the Fermi sea and far above the Fermi 
momentum 
the momentum dependence nearly saturates. For the $\sigma\omega$-potential
the self-energies are rather weakly momentum dependent: 
over the full interval they decrease by about 20\%. For comparison we also
show two results obtained by the Groningen group.
In Ref. \cite{6} the self-energies above the Fermi surface are calculated
in a similar way as in the present work, see discussion of Fig.\,5.
The results obtained by the Groningen group \cite{6} are in general softer 
k-dependent than those of the present work obtained with the Bonn potentials. 
The results of Ref. \cite{9} show
a momentum dependence as strongly as ours.
However, in this case a comparison is not fully appropriate
since these calculations include an additional 
self-consistent treatment of the pion polarisation.
\subsection{equation of state and optical potential}
In Fig.\,8 we show the equation of state, i.e., the energy per particle 
E/A as a function of the nuclear matter density $\varrho$, 
calculated with Bonn A,B,C. 
In the relativistic Brueckner theory the energy per particle is defined 
in analogy to the non-relativistic Hartree-Fock method as the kinetic 
plus half of the potential energy
\begin{eqnarray}                                      \label{eq26}
& &  {\rm E / A }   =   \frac{1}{\varrho} \; \sum_{{\bf k},\lambda} 
< \overline{u}_{\lambda}({\bf k})|
\mbox{\boldmath$ \gamma $ \unboldmath} \cdot {\bf k} + M +
\frac{1}{2}\Sigma(k)|u_{\lambda}({\bf k}) > \frac{M^{\ast}}{E^{\ast}} - M    
                                                          \\ \label{eq27} 
& &  =  \frac{1}{\varrho} \int_F \frac{d^3{\bf k}}{2\pi^3} 
    \left[ ( (1 + \Sigma_v(|{\bf k}|))E^{\ast} - \Sigma^0(|{\bf k}|) ) 
-\frac{1}{2E^{\ast}}( \Sigma_s(|{\bf k}|) M^{\ast} - \Sigma_{\mu}(|{\bf k}|) 
k^{\ast\mu} ) \right] - M 
\end{eqnarray}
%\begin{eqnarray}                                      \label{eq26}
%&& {\rm E / A }   =   \frac{1}{\varrho} \; \sum_{{\bf k},\lambda} 
%< \overline{u}_{\lambda}({\bf k})|
%\mbox{\boldmath$ \gamma $ \unboldmath} \cdot {\bf k} + M +
%\frac{1}{2}\Sigma(k)|u_{\lambda}({\bf k}) > \frac{M^{\ast}}{E^{\ast}} - M    
%                                                        \\ \label{eq27} 
%&&  =  \frac{1}{\varrho} \int_F \frac{d^3{\bf k}}{2\pi^3} 
%    \left[ ( (1 + \Sigma_v(k))E^{\ast} - \Sigma^0(k) ) 
%-\frac{1}{2}( \Sigma_s(k) M^{\ast} - \Sigma_{\mu}(k) k^{\ast\mu} ) / E^{\ast}
%\right] - M 
%\end{eqnarray}
with the self-consistent spinors $|u_{\lambda}({\bf k}) >$ given in 
Eq. (\ref{eq9}).
By this way we have determined E/A by integrating the self-energy components
of Eqs. (\ref{eq16}-\ref{eq18}) over the Fermi sphere, see Eq. (\ref{eq27}). 
In Fig.\,8 we compare to the results of Ref. \cite{5} where E/A is 
calculated directly from the single particle potential, Eq. (\ref{eq24}),
using the expression (\ref{eq26}). For moderate densities 
the results agree with an accuracy better 0.5 MeV and 
even for $k_F=1.9 fm^{-1}$ 
the deviation is less than 3 MeV. Firstly, this result is an additional
test for the accuracy of our numerical treatment, i.e., the projection of the 
T-matrix onto covariant amplitudes which are subsequently used to determine 
the self-energy components and the equation of state. 
Secondly, these results again demonstrate that the equation of state is quite 
insensitive to the self-energy components themselves. 
Actually, as can be seen clearly from the definition Eq. (\ref{eq26}), 
the equation of state depends directly on the single-particle potential, 
i.e., the difference of $\Sigma_s$ and $\Sigma_0$, Eq. (\ref{eq25}), 
and relatively weak on the effective mass via the kinetic energy part.
Nevertheless, the relativistic effects are responsible in contrast to 
non-relativistic calculations that the minima of the 
equations of state for the three different
Bonn potentials reveal a Coester band which meets the empirically 
found saturation area of nuclear matter.

In order to demonstrate the influence of the two different choices, i.e.,
pseudo-vector and pseudo-scalar on the equation of state
in Fig.\,9 we show the respective equation of state, again for Bonn C.
Actually the expectation value
$<{\overline u}| \Sigma | u>$ of the full self-energy operator,
Eq. (\ref{eq16a}), is identical for both choices \cite{19}.
However, as can be seen from Fig.\,9 the equation of state
turns out to be rather sensitive on the different choices.
This significant dependence is only due to 
the self-consistently iterated effective mass which differs in
both approaches.
This effect deepens the equation of state in the pseudo-scalar choice 
by around 1 to 2 MeV and shifts the minimum towards higher densities.
Thus the saturation point is shifted away from the empirical point
towards the results of non-relativistic Brueckner calculations
as, e.g., published in Ref. \cite{5}. This observation additionally
supports the pseudo-vector choice.

In view of the strong momentum dependence of the self-energy components
for the Bonn potentials one might expect that the results depend
sensitive on the actual momentum reference value  where the selfconsistent
effective mass is determined. As a consequence a selfconsistent treatment 
of the full momentum dependence might appear to be indispensable. Therefore
we have also examined the influence of a different
choice of the selfconsistent effective mass.
In the reference spectrum approximation, see Eq. (\ref{eq10}), $M^{\ast}$ 
is determined at the Fermi surface which is the conventional treatment in the 
relativistic Brueckner approach. However, this choice is by no means unique.
Thus we also determine $M^{\ast}$ 
by its mean value in analogy to Eq. (\ref{eq10})
\begin{equation}                                    \label{eq28}
	M^{\ast} = M + {\overline \Sigma_s}(M^{\ast}) 
	- M^{\ast} {\overline \Sigma_v}(M^{\ast}) .
\end{equation}
The self-energies $\Sigma_{s,v}$ are not simply taken 
at $|{\bf k}|=k_F$ but replaced by their values
averaged over the Fermi sea
\begin{equation}                                       \label{eq29}
	\overline{\Sigma}_{s,v} =  \frac{ 
	\int \Sigma_{s,v}(|{\bf k}| ) \Theta(k_F-|{\bf k}|)
	M^{\ast} / E^{\ast}  d^3 {\bf k}   }
	{     \int  \Theta(k_F-|{\bf k}|)
	 M^{\ast} / E^{\ast}  d^3 {\bf k}   } .
\end{equation}
These averages are Lorentz scalars.
The extension of the reference spectrum approximation, Eq. (\ref{eq28}), 
is especially meaningful with respect to the treatment of 
anisotropic nuclear matter, as done in Ref. \cite{13}. 
Here it serves as a test of the influence 
of the momentum dependence. Due to the strong momentum 
dependence observed, e.g., in Fig.\,7 the mean value ${\overline \Sigma_s}$
differs considerably from $\Sigma_s(k_F)$, i.e., 
by about 57 MeV at saturation density (with Bonn C).
Respectively for the different effective masses we obtain a 
mean value $M^{\ast}=528$ MeV which is smaller than 
the selfconsistent mass $M^{\ast}(k_F)=566$ MeV.
But if we use the alternatively defined effective mass, Eq. (\ref{eq28}),
as iteration parameter we find for the  
final iterated mean value $ M^{\ast}=547$ MeV.
Thus the mean value is increased, i.e., it is shifted towards the previous
self-consistent result or, in other words, the selfconsistent value 
is rather stable.
The comparison of the self-energy components $\Sigma_{s,0}(k_F)$
in both iterations, see Fig.\,2, shows  just slightly deviations, 
e.g., about 10 MeV at saturation density.
And even more, the equation of state, shown in Fig.\,9, remains 
nearly unchanged.
These results demonstrate that the strong momentum 
dependence has only minor influence on the density dependence
of the self-energy components and that the equation of state remains 
unaffected.

Another quantity which is more sensitive on the momentum dependence of 
$\Sigma_{s,0,v}$ is 
the Schroedinger equivalent optical potential which a nucleon feels inside 
the nuclear medium. The optical potential is given by
\begin{equation}                                      \label{eq30}
U_{\rm opt}(|{\bf k}|,k^0) 
=  \Sigma_s(|{\bf k}|) - \frac{1}{M} k^{\mu} \Sigma_{\mu}(|{\bf k}|) 
	+ \frac{\Sigma_s^2(|{\bf k}|) - \Sigma_{\mu}^2(|{\bf k}|)}{2M}  ~.
\end{equation}
It contains counteracting  linear and quadratic terms in the
self-energy components. For nucleons with $|{\bf k}|>k_F$ 
the momentum space is not 
completely blocked for scattering processes and thus
above the Fermi surface the self-energies and the optical potential 
become complex. The single-particle energy 
$k^0(|{\bf k}|)=-\Sigma^0+E^{\ast}(|{\bf k}|)$ is a monotonic function 
of $|{\bf k}|$ which allows by inversion to determine $U_{\rm opt}(k^0)$
as function of the single-particle energy $k^0$. 
We want to stress that neglecting
the momentum dependence of the fields $\Sigma_{s,(0)}$ and 
neglecting completely $\Sigma_v$ the optical potential $U_{\rm opt}$ 
would simply be a linear function of $k^0$.
However, this is not the case here.
Real and imaginary
part of the optical potential calculated for Bonn C (pseudo-vector choice)
are shown in Fig.\,10. The real part agrees well with the Groningen results 
of Ref. \cite{7}  while the imaginary part of the present calculations 
is about twice as 
large. Further we compare to the empirical optical potential
Hama I, table 2 of Ref. \cite{21}. For the full range of single-particle
energies up to 700 MeV, shown in Fig.\,10, our results deviate from the 
empirical Re $U_{\rm opt}$ by less than 10 MeV. Up to the pion 
production threshold at around 300 MeV we observe a remarkable
agreement of the imaginary part of $U_{\rm opt}$
with the empirical data. For single-particle energies above this region 
the results appear to be no more fully reliable since then 
meson-nucleon resonances, e.g. the $\Delta$(1232) resonance, 
should be taken into account. 

\section{Summary}
In this work in particular the self-energy of a nucleon inside the 
nuclear medium
has been studied in the framework of the relativistic Brueckner approach.
Therefore the dependence of the Lorentz scalar and vector 
self-energy components 
on the nuclear density and the nucleon momentum have been examined.
As bare nucleon-nucleon interaction various one-boson exchange potentials
have been employed, i.e., the Bonn A, B, C potentials and for comparison the 
$\sigma\omega$-exchange potential. For the latter we are able to 
nicely reproduce 
the density dependence of the self-energies of Ref. \cite{1}.
The self-energy components themselves deviate for the three different 
Bonn potentials over a wide density range only by a few MeV  and thus 
we restricted our discussion mainly to the Bonn C potential.
From the comparison with our results we find that 
the potential-fit method of Ref. \cite{5} underpredicts the magnitude 
of the self-energies by about 50 up to more than 150 MeV.

Further we analyzed the influence of different possible covariant 
representations of the T-matrix. 
The positive energy on-shell T-matrix is projected on 
five invariant amplitudes and by the restriction to positive energy states
an ambiguity arises which is discussed in terms of the pseudo-scalar 
and the pseudo-vector representation. 
Although the T-amplitudes of both choices are equal 
on the one-body level, i.e., concerning the respective self-energy 
components a significant influence of these representations is observed.
This is due to the fact that
in the pseudo-scalar case direct and exchange contributions 
to the self-energies are equal
while in the  pseudo-vector choice the exchange part is suppressed.
For the Bonn potentials this suppression lies between 40 - 80 MeV
and is most prominant at low densities.
This difference between both choices has to be interpreted as an
inherent measure for an uncertainty of the approach due to the restriction 
on positive energy spinors. 
  
Commonly the momentum dependence of the self-energy components is supposed
to be weak. This assumption is also reflected in the self-consistency 
treatment in the relativistic Brueckner model which is managed by just 
one density dependent parameter, i.e., the self-consistent effective mass.
Therefore we have carefully examined this momentum dependence which, however, 
can be done within the Bruckner scheme just in a first approximation.
As in Ref. \cite{1} we find that taking the $\sigma\omega$-exchange 
potential the self-energies are found 
to be constant below the Fermi surface within a few percent.
However, this result does not hold for the more realistic Bonn potentials.
Here we observe a strong momentum dependence which may be due to the 
strong pion exchange. 
The self-energies decrease from
their central value, i.e., at $k=0$ to the Fermi surface by about 
one fourth and further to $k=0.8$ GeV by more than one half. 
This striking momentum dependence makes the treatment of the 
self-consistency in the model somewhat questionable. But a test with
a different defined self-consistent effective mass 
which is determined by a mean value over the Fermi sea
and not taken at $k_F$ as usually done
shows that the self-energy components are rather stable at not too high
densities. Thus the self-consistency treatment in the model 
still appears to be justified up to two or three times saturation density. 

In this context  the question raises
which observables are sensitive to the self-energy components 
and their momentum dependence.
It is remarkable that the full self-energy and, respectively, 
the single particle potential show (absolutely) a much weaker
momentum dependence.
As it is well known from mean field theory the large scalar and vector 
self-energy components contribute to these quantities with opposite signs 
and thus cancel each other to hight extent.
The resulting observables, e.g., the single particle potential 
are smaller by nearly one order of magnitude. 
Here we found that this holds not only at $k_F$ but also over
the full momentum range considered. 
Thus the momentum dependence cancels to a large extent. 
Also the equation of state depends mainly
on the difference of the scalar and the vector self-energy and 
thus the absolute values of the single component are of minor importance. 
This can be seen most clearly by the use of 
just density dependent self-energy components determined with the 
potential-fit method of Brockmann and Machleidt  \cite{5} 
which deviate from our results (at $k_F$) by about 75 MeV. 
The resulting binding energy, however, only deviates  
by about 1 MeV.
Similarly the self-energy components obtained in the
pseudo-scalar and the pseudo-vector choice differ considerable but the
equations of state differ by less than one to two MeV.
On the other hand,  small variations of the self-energy 
due to the different three Bonn potentials 
produce a significant change in the equations of state.
Summarizing, the  equation of state is not sensitive to the 
self-energy components themselves but to their difference.
A bit more sensitive is another quantity, i.e.,  
the optical potential. It depends not only on the cancellation
of the scalar and vector self-energy 
but also contains quadratic terms. We find a good agreement of our
results with the experimental data of Ref. \cite{21} 
for, both, the real and the imaginary part 
up to a single particle energy of several hundred MeV.

As a resum\'e, the self-energy components in the 
relativistic Brueckner model are still affected with some uncertainty.
For realistic one-boson exchange potentials we find them to be 
strongly momentum dependent. However, for this attribute we have to be aware 
of the limits of the model.
One major success of the relativistic approach was that the
equation of state comes close to the empirical saturation point.
The equation of state is rather insensitive to the 
self-energy components themselves. 
Similarly, the nucleon optical potential is only little more sensitive.
However, a precise knowledge of the size of each self-energy 
component is necessary, e.g., in the much more complicate situation of 
heavy ion collisions.

\acknowledgements
The authors acknowledge stimulating discussions with 
F. de Jong and H. M\"uther.

%%%%%%%%%%%%%%%%%%%%%%%%%%%%%%%%%%%%%%%%%%%%%%%%%%%%%%%%%%%%%%%%%%%%%%%%%%%%

%%%%%%%%%%%%%%%%%%%%%%%%%%%%%%%%%%%%%%%%%%%%%%%%%%%%%%%%%%%%%%%%%%%%%%%%%%%%
        \newpage
	\section*{Figure Captions}
	\begin{itemize}
	\item[Fig.\,1:] Self-consistent effective mass $M^{\ast}$ 
        in units of the bare mass
        ($M$=939 MeV) as a function of the Fermi momentum $k_F$. The 
        relativistic Brueckner results obtained with the 
        $\sigma\omega$-exchange
        potential (full line) are compared to those of
        Ref. \protect\cite{1} (diamonds). The effective mass for the Bonn C 
        potential (dashed line) differs significantly from that of the
        $\sigma\omega$-model.
     
	\item[Fig.\,2:] Self-energy components for the pseudo-vector choice, 
        calculated with various Bonn potentials.
        The dotted, dashed and full lines represent the Bonn A,B and C
        potential. The scalar self-energy
        $\Sigma_s$ (lower half-plane) and the time- and space-like vector 
        self-energies $-\Sigma_0,- k_F \Sigma_v$ (upper half-plane) 
        are shown as functions of $k_F$. For the three Bonn potentials 
        the small space-like vector self-energies $- k_F \Sigma_v$ 
        are equal within line width.
        The remaining curves (dashed dotted) involve a different definition 
        of the 
        self-consistent effective mass which is given in the text later on. 

	\item[Fig.\,3:] Self-energy components $\Sigma_s,-\Sigma_0$  
        calculated with the pseudo-scalar choice and the Bonn C potential.
        The curves 
        represent the total self-energy components (full lines), their
        direct (dir.) and their exchange (exc.) parts 
        which are practically equal (dashed dotted lines).

	\item[Fig.\,4:] Self-energy components $\Sigma_s,-\Sigma_0$  
        calculated with the pseudo-vector choice and the Bonn C potential.
        The curves 
        represent the total self-energy components (full lines), their
        direct (dir.) parts (dashed lines) and their exchange (exc.) 
        parts (dotted lines).

	\item[Fig.\,5:] Self-energy components $\Sigma_s,-\Sigma_0$  
        calculated with the pseudo-vector choice and the Bonn C potential
        (full lines). They are compared with the 
        results of Ref. \protect\cite{5} (dashed lines) and of 
        Ref. \protect\cite{6} (dotted 
        lines).

	\item[Fig.\,6:] Self-energy components 
        $\Sigma_s,-\Sigma_0$ at saturation 
        density $(k_F=0.28$ GeV) as function of the nucleon 
        momentum $|{\bf k}|<k_F$.
        The self-energies are calculated with the 
        $\sigma\omega$-exchange potential (pseudo-scalar choice) and 
        compared with values (symbols) 
        taken from Ref. \protect\cite{1}.

	\item[Fig.\,7:] Self-energy components $\Sigma_s,-\Sigma_0$  
        as function of the momentum $|{\bf k}|$ 
        of the nucleon for $k_F=0.265$ GeV. 
        The self-energies are calculated with the 
        Bonn C (pseudo-vector choice, solid line) 
        and the $\sigma\omega$-exchange potential 
        (pseudo-scalar choice, dotted line) 
        and compared with results of Ref. \protect\cite{6} (dashed line) 
        and of Ref. \protect\cite{9} ($k_F=0.27$ GeV, dashed dotted line).

	\item[Fig.\,8:] Equation of state, i.e., energy per particle $E/A$ 
        as a function of the nuclear matter density $\varrho$, 
        for the three Bonn potentials A,B and C
        (from below to top: dashed line, dotted line and solid line).
        The symbols represent the respective results of 
        Ref. \protect\cite{5}.

	\item[Fig.\,9:] Equation of state for the Bonn C potential calculated
        in the pseudo-vector (solid line) and the pseudo-scalar
        (dashed line) choice.
        The remaining curve (dotted) involves a different definition 
        of the 
        self-consistent effective mass given in the text.

        \item[Fig.\,10:] Real and imaginary part of the 
        Schroedinger equivalent optical potential $U_{\rm opt}$ 
        as function of the single-paricle energy $k^0 - M$.
        The results for the Bonn C potential 
        (real part: solid line, imaginary part: dotted line)
        are compared to the results of Ref. \protect\cite{6}
        (real part: dashed line, imaginary part: dashed dotted line)
        and to the empirical optical potential taken from Ref. 
        \protect\cite{21} (real part: diamonds, imaginary part: circles).

	\end{itemize}
%%%%%%%%%%%%%%%%%%%%%%%%%%%%%%%%%%%%%%%%%%%%%%%%%%%%%%%%%%%%%%%%%%%

        \begin{figure}
        \begin{center}
        \leavevmode
        \epsfxsize = 17.5cm
        \epsffile[28  65  540  588]{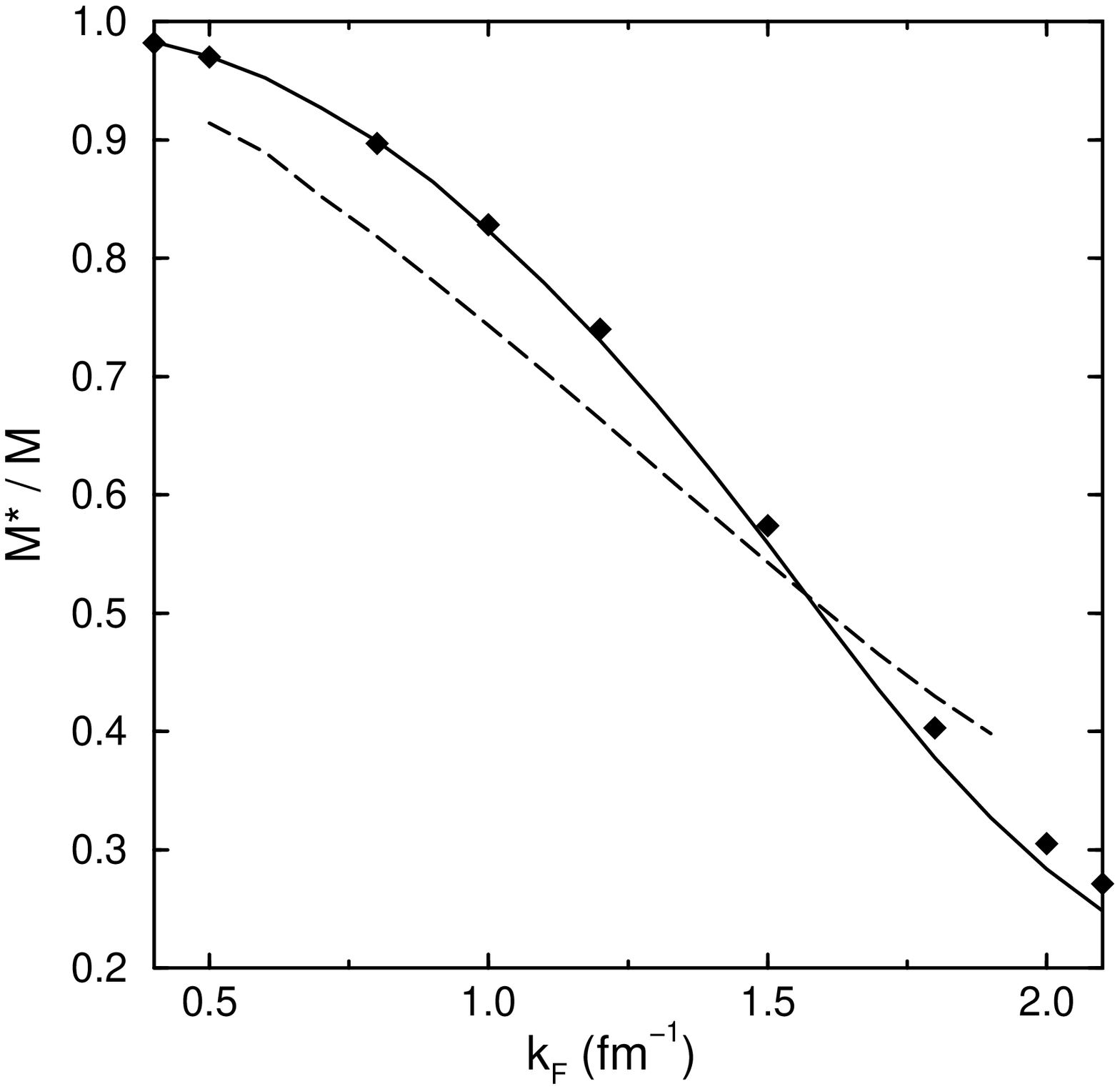}
        \end{center}
        \caption{}
        \label{fig1}
        \end{figure}

        \begin{figure}
        \begin{center}
        \leavevmode
        \epsfxsize = 17.5cm
        \epsffile[28  65  540  588]{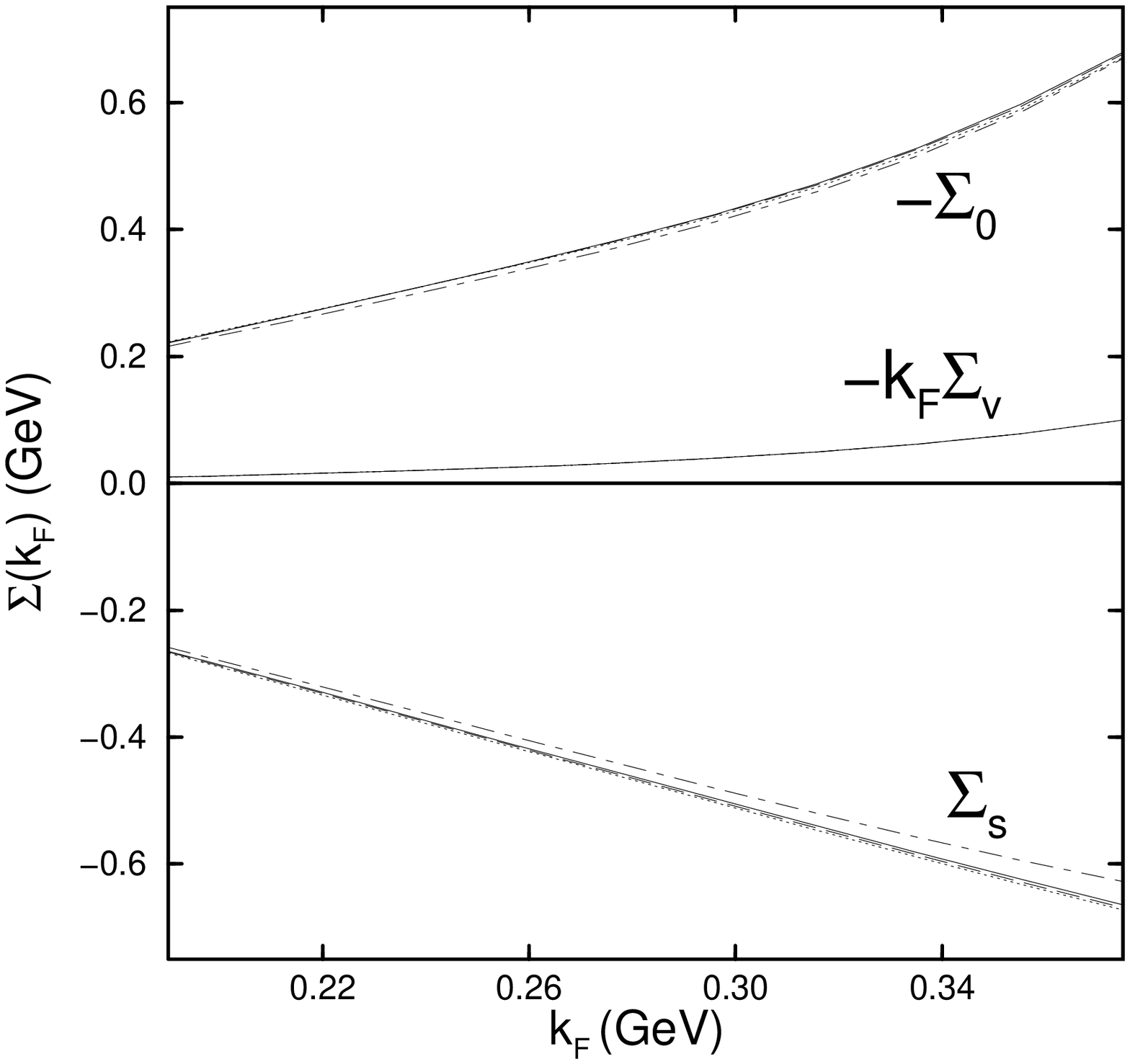}
        \end{center}
        \caption{}
        \label{fig2}
        \end{figure}

        \begin{figure}
        \begin{center}
        \leavevmode
        \epsfxsize = 17.5cm
        \epsffile[28  65  540  588]{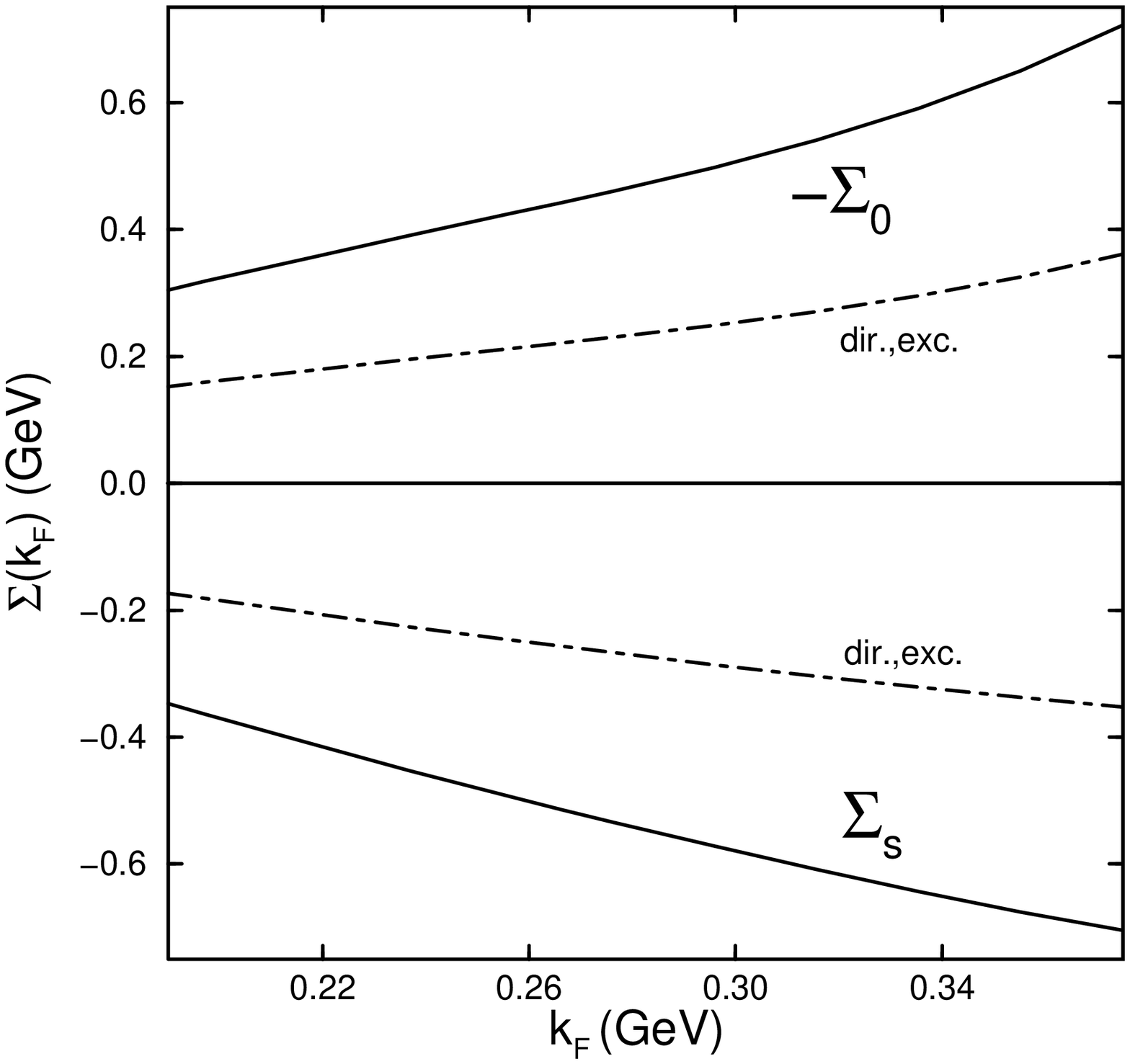}
        \end{center}
        \caption{}
        \label{fig3}
        \end{figure}

        \begin{figure}
        \begin{center}
        \leavevmode
        \epsfxsize = 17.5cm
        \epsffile[28  65  540  588]{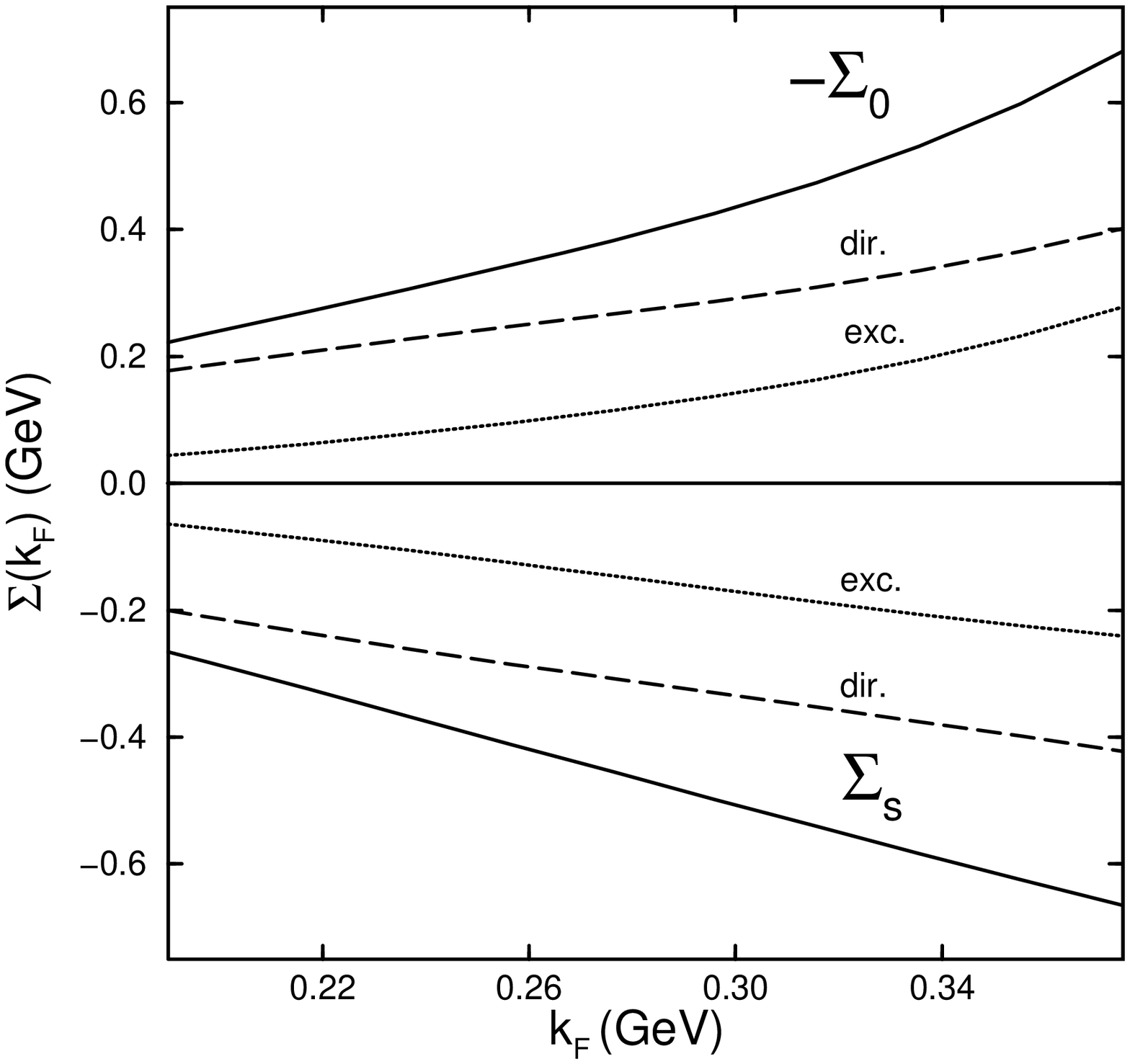}
        \end{center}
        \caption{}
        \label{fig4}
        \end{figure}

        \begin{figure}
        \begin{center}
        \leavevmode
        \epsfxsize = 17.5cm
        \epsffile[28  65  540  588]{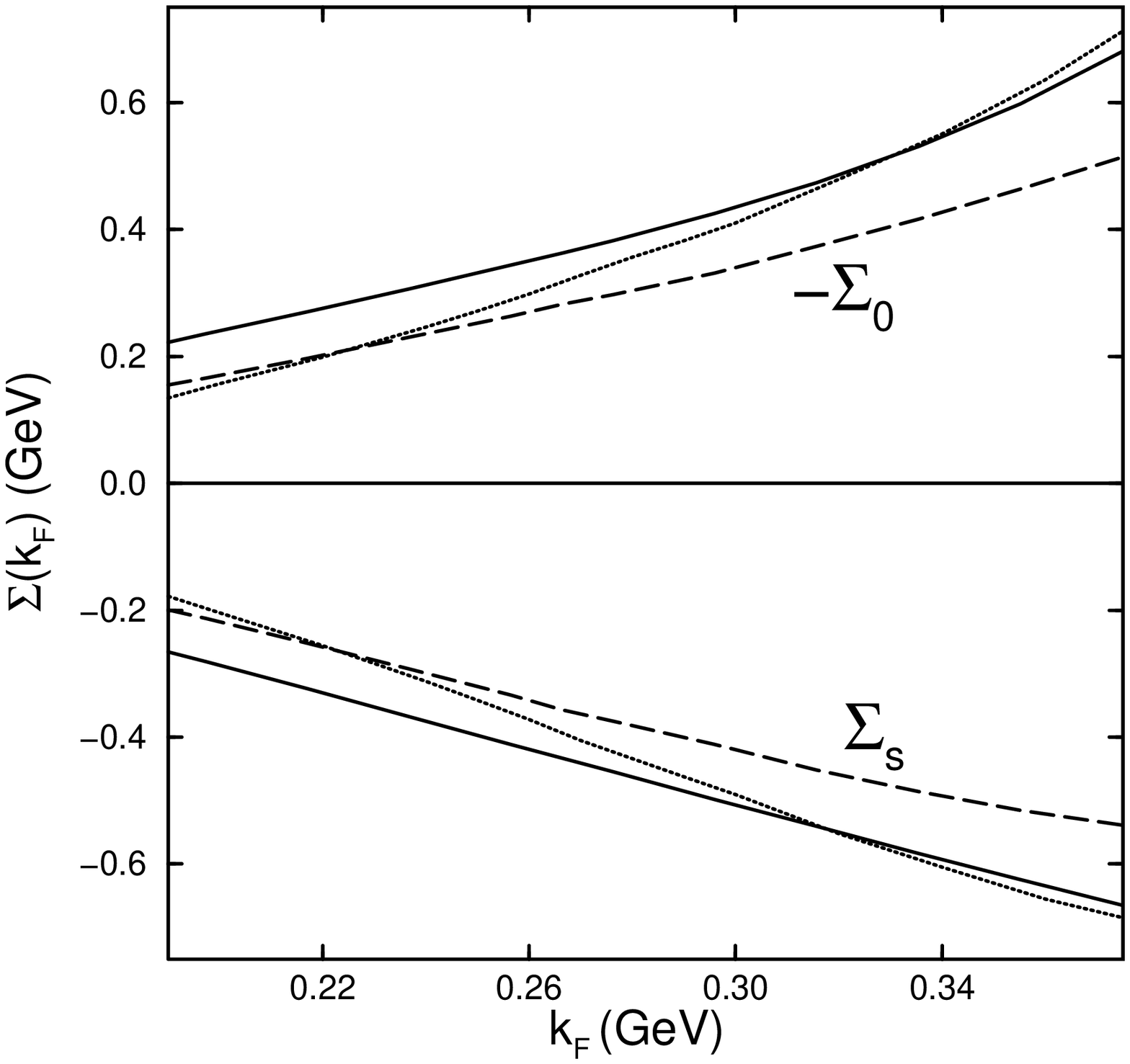}
        \end{center}
        \caption{}
        \label{fig5}
        \end{figure}

        \begin{figure}
        \begin{center}
        \leavevmode
        \epsfxsize = 17.5cm
        \epsffile[28  65  540  588]{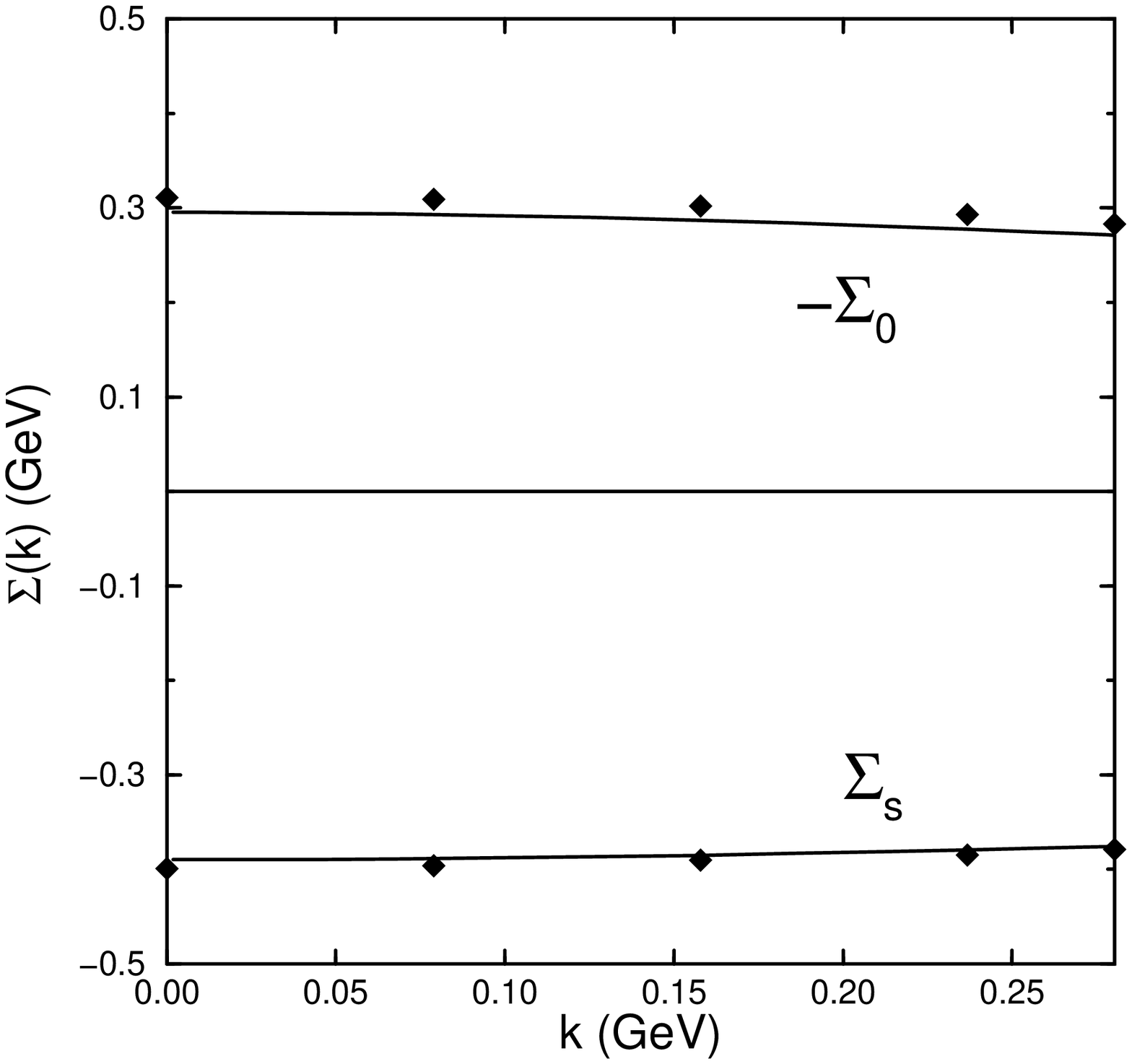}
        \end{center}
        \caption{}
        \label{fig6}
        \end{figure}

        \begin{figure}
        \begin{center}
        \leavevmode
        \epsfxsize = 17.5cm
        \epsffile[28  65  540  588]{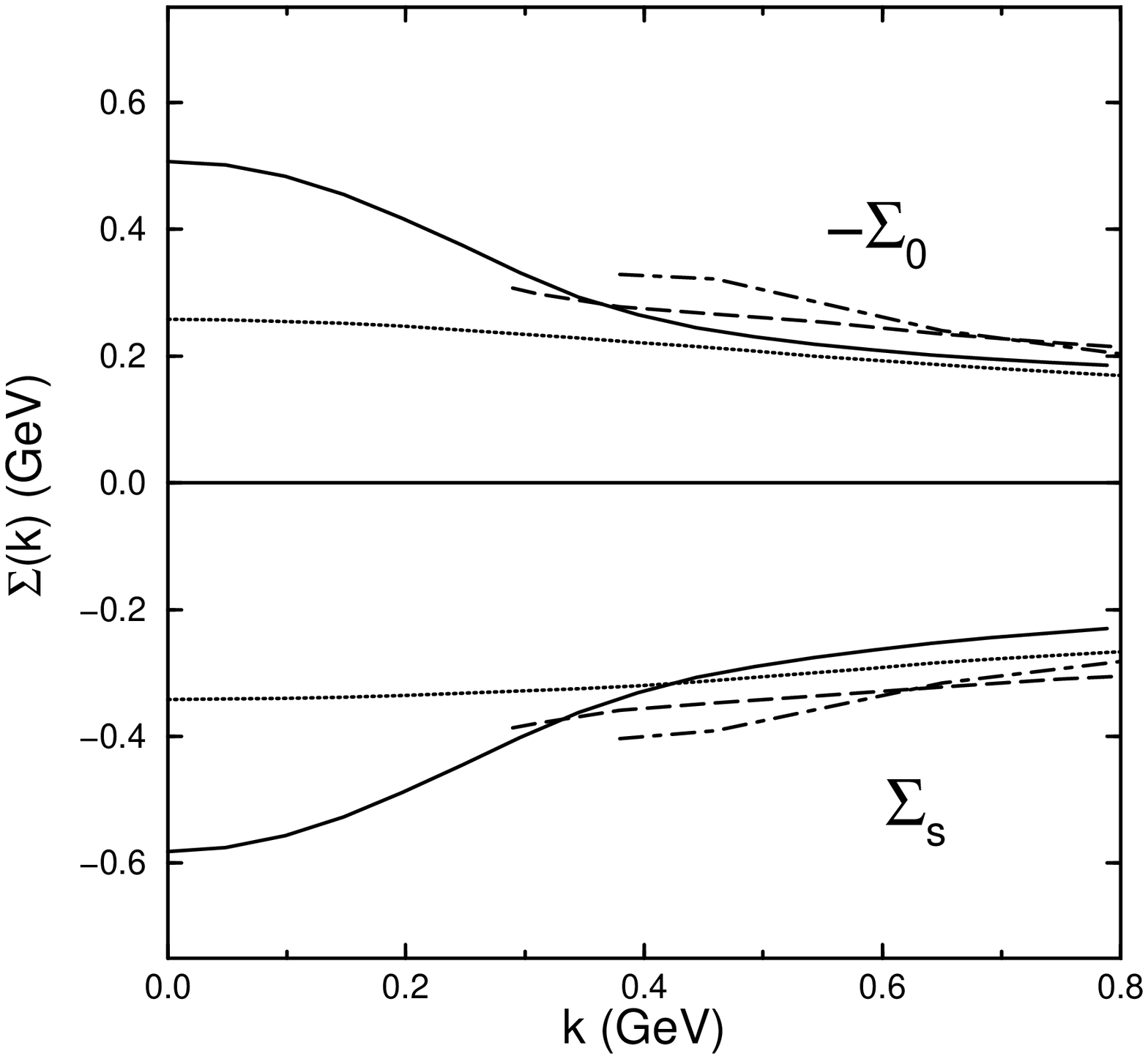}
        \end{center}
        \caption{}
        \label{fig7}
        \end{figure}

        \begin{figure}
        \begin{center}
        \leavevmode
        \epsfxsize = 17.5cm
        \epsffile[28  65  540  588]{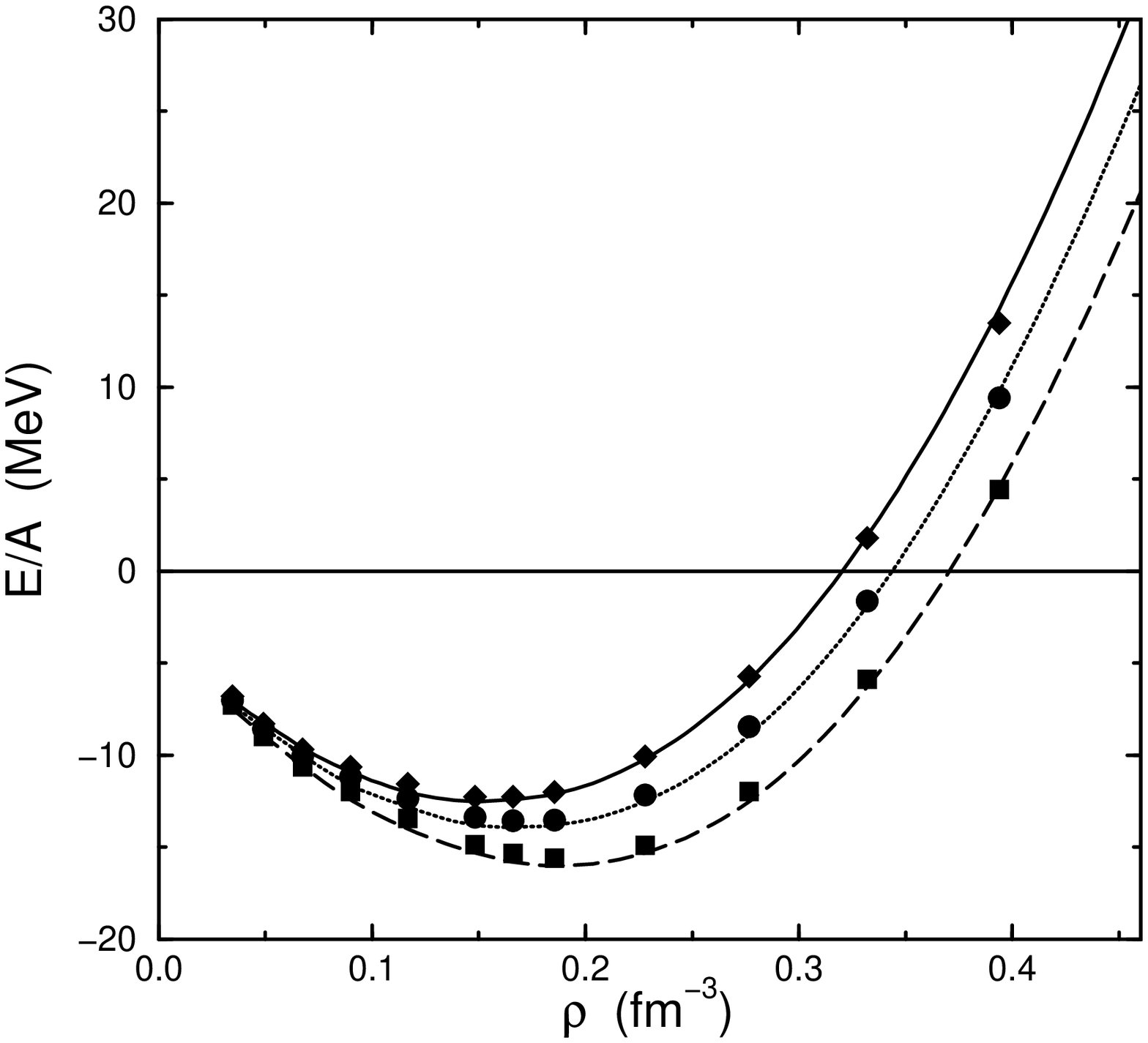}
        \end{center}
        \caption{}
        \label{fig8}
        \end{figure}

        \begin{figure}
        \begin{center}
        \leavevmode
        \epsfxsize = 17.5cm
        \epsffile[28  65  540  588]{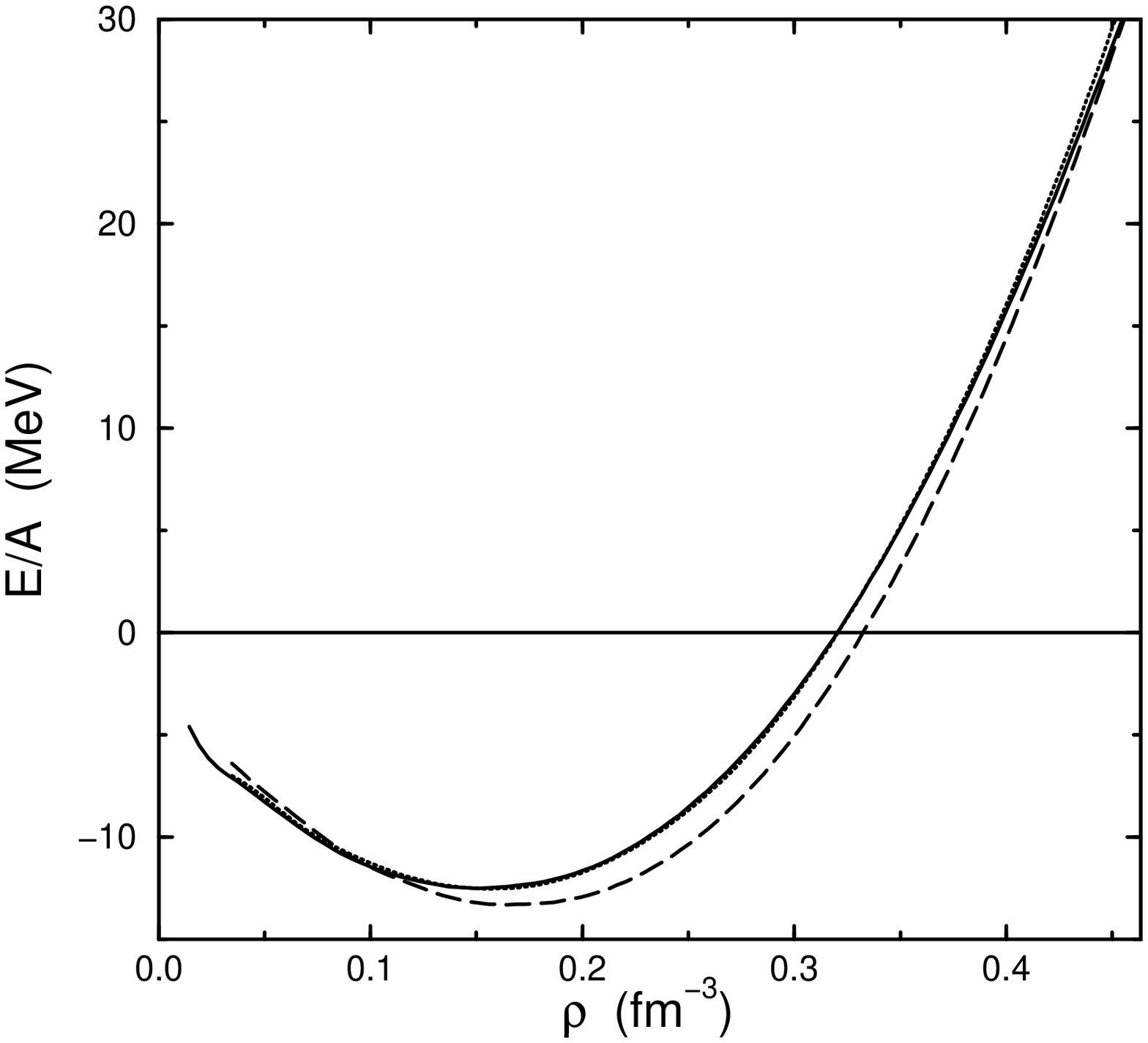}
        \end{center}
        \caption{}
        \label{fig9}
        \end{figure}

        \begin{figure}
        \begin{center}
        \leavevmode
        \epsfxsize = 17.5cm
        \epsffile[28  65  540  588]{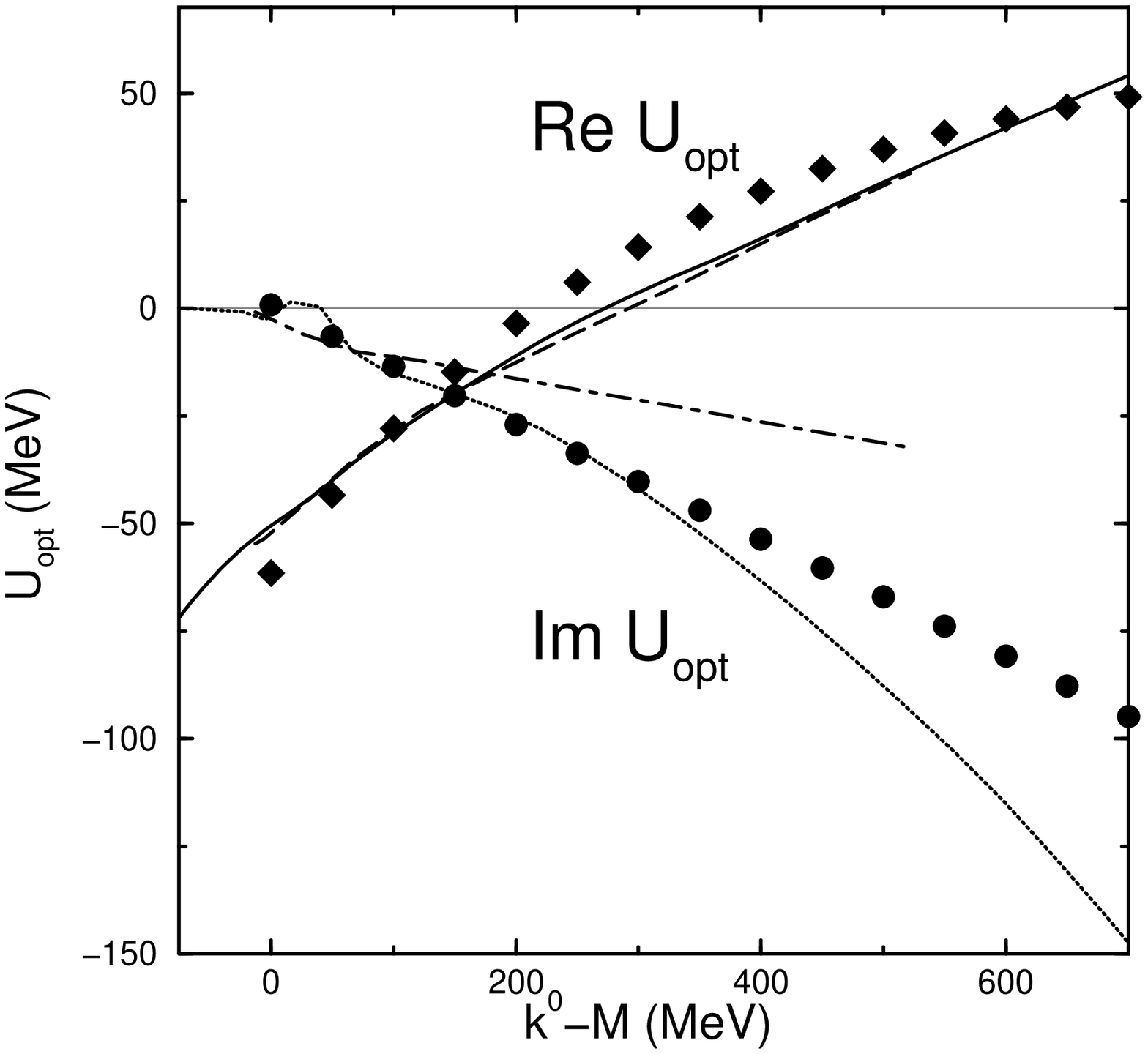}
        \end{center}
        \caption{}
        \label{fig10}
        \end{figure}

%%%%%%%%%%%%%%%%%%%%%%%%%%%%%%%%%%%%%%%%%%%%%%%%%%%%%%%%%%%%%%%%%%%%
\end{document}